\documentclass[12pt,letter]{article}
\pdfoutput=1
\usepackage{graphicx, epsfig, color,cite,inputenc}
\usepackage{amsmath}
\usepackage{amssymb}
\usepackage{float}
\usepackage{caption,subcaption,graphicx}
\usepackage{hyperref}
\usepackage{slashed}
\def\to{\rightarrow}

\def\bi{\begin{itemize}}
\def\ei{\end{itemize}}

\def\ta{\tilde a}
\def\tchi{\tilde\chi}
\def\dsl{\partial \llap/}

\def\tst{\tilde t}

\def\tg{\tilde g}

\def\alt{\lesssim}
\def\agt{\gtrsim}
\def\be{\begin{equation}}  
\def\ee{\end{equation}}  
\def\bea{\begin{eqnarray}}  
\def\eea{\end{eqnarray}}

\begin{document}
\begin{titlepage}
\begin{flushright}
OU-HEP-230204
\end{flushright}

\vspace{0.5cm}
\begin{center}
  {\Large \bf Dark matter and dark radiation from the\\
    early universe with a modulus coupled to the PQMSSM
}\\ 
\vspace{1.2cm} \renewcommand{\thefootnote}{\fnsymbol{footnote}}
{\large Howard Baer$^{1}$\footnote[1]{Email: baer@ou.edu },
Vernon Barger$^2$\footnote[2]{Email: barger@pheno.wisc.edu} and
Robert Wiley Deal$^{1}$\footnote[3]{Email: rwileydeal@ou.edu} 
}\\ 
\vspace{1.2cm} \renewcommand{\thefootnote}{\arabic{footnote}}
{\it 
$^1$Homer L. Dodge Department of Physics and Astronomy,
University of Oklahoma, Norman, OK 73019, USA \\[3pt]
}
{\it 
$^2$Department of Physics,
University of Wisconsin, Madison, WI 53706 USA \\[3pt]
}

\end{center}

\vspace{0.5cm}
\begin{abstract}
\noindent
The supersymmetrized DFSZ axion model is especially compelling in that it
contains 1. the SUSY solution to the gauge hierarchy problem, 2. the
Peccei-Quinn (PQ) solution to the strong CP problem and 3. the Kim-Nilles
solution to the SUSY $\mu$ problem.
In a string setting, where a discrete $R$-symmetry
(${\bf Z}_{24}^R$ for example) may emerge from the compactification
process, a high-quality accidental axion (accion) can emerge from the
accidental, approximate remnant global $U(1)_{PQ}$ symmetry where the
decay constant $f_a$ is linked to the SUSY breaking scale, and is within
the cosmological sweet zone. In this setup, one also expects the presence of
stringy remnant moduli fields $\phi_i$. Here, we consider the situation of a
single light modulus $\phi$ coupled to the PQMSSM in the early universe,
with mixed axion plus higgsino-like WIMP dark matter.
We evaluate dark matter and dark radiation production via nine coupled
Boltzmann equations and assess the severity of the cosmological moduli
problem (CMP) along with dark matter and dark radiation production rates.
We find that typically the light modulus mass should be $m_{\phi}\agt 10^4$
TeV to avoid the moduli-induced dark matter overproduction problem.
If one is able to (anthropically) tune the modulus field amplitude,
we find a value of $\phi_0\alt 10^{-7}m_P$ would be required to solve the overall CMP.
\end{abstract}
\end{titlepage}

\section{Introduction}
\label{sec:intro}

The Standard Model (SM), for all its successes, is beset with problems:
among them
1. the gauge hierarchy problem (where quantum corrections drive the Higgs mass
$m_h$ up to the highest mass scales for which the SM is valid),
2. the strong CP problem (why is the QCD $\bar{\theta}$ parameter so tiny, $\alt 10^{-10}$) and 
3. the inclusion of gravity into the model.
In addition, the SM provides no explanation for the dark matter/dark energy
in the universe, the matter-antimatter asymmetry or the source field for
inflation (inflaton).
In this paper, we wish to explore some of the phenomenological/cosmological
consequences of what might be considered the
most plausible extension of the SM that solves these various issues:
the supersymmetrized SM (Minimal Supersymmetric Standard Model or MSSM\cite{Baer:2006rs}),
coupled with a Peccei-Quinn axion sector where we adopt the supersymmetrized
DFSZ\cite{Dine:1981rt,Zhitnitsky:1980tq} axion which fits snugly into the MSSM framework since
both setups necessarily include the required two-Higgs doublets,
and the SUSY DFSZ setup\cite{Bae:2011iw,Bae:2013bva,Bae:2013hma} contains the elegant Kim-Nilles\cite{Kim:1983dt} solution to the
SUSY $\mu$ problem\cite{Bae:2019dgg}. The $R$-parity conserving MSSM contains a
WIMP dark matter candidate\cite{Goldberg:1983nd,Ellis:1983ew}, the lightest neutralino $\tchi$, while the
DFSZ model contains the QCD axion $a$, also a dark matter candidate\cite{Abbott:1982af,Preskill:1982cy,Dine:1982ah}. Thus,
the model to be considered here contains {\it two} dark matter candidates,
both $\tchi$ and $a$\cite{Baer:2014eja}.

The structure of the SUSY DFSZ model has been elucidated in several
previous papers\cite{Bae:2011iw,Bae:2013bva,Bae:2013hma}.
For the SUSY DFSZ model, the relic abundance isn't as simple as just
summing the usual thermally-produced WIMPs plus the coherent-oscillation (CO)
produced axions\cite{Abbott:1982af,Preskill:1982cy,Dine:1982ah} (for the case where PQ symmetry is broken before the
end of inflation, which we restrict ourselves to in this paper).
Instead, one must also factor in the presence of thermally produced
axinos $\ta$, the spin-1/2 partners of the axions\cite{Choi:2008zq,Baer:2011hx}. Once produced in the early
universe, they can cascade decay into LSPs and thus add to the WIMP abundance.
Also, one must consider the spin-0 scalar axion partners, the saxions $s$.
Saxions can be produced both thermally and non-thermally via COs, and
undergo delayed decays to SM particles (leading to entropy dilution of
any relics present at the time of decay) as well as decays to SUSY particles
(thus adding a non-thermal component to the WIMP abundance)\cite{Baer:2011uz}, and they can
decay to axions $s\to aa$ leading to dark radiation\cite{Bae:2013qr}, which is constrained
by limits on the effective number of additional neutrinos in the early
universe, $\Delta N_{eff}$. The Planck 2018\cite{Planck:2018vyg} averaged limit
finds $N_{eff} =2.99\pm 0.17$ which may be compared to the SM
expectation that $N_{eff}^{SM}=3.046$ so that $\Delta N_{eff}< 0.29$ at 95\% CL.
Along with production of axinos and saxions, it is proper to include
the presence of the spin-3/2 gravitinos $\psi_\mu$\cite{Kawasaki:1994af,Kohri:2005wn,Kawasaki:2008qe}.
Gravitinos may be
produced thermally at large rates in the early universe, depending
on $T_R$, the temperature of radiation after inflaton decay\cite{Pradler:2006hh}.
They can then cascade decay to $\tchi$ states, thus potentially
adding to the non-thermal presence of WIMP dark matter or disrupting successful Big Bang Nucleosynthesis (BBN).

Dark matter production in the SUSY DFSZ model-- including MSSM fields plus
axions, saxions, axinos and gravitinos-- requires solving eight coupled
Boltzmann equations which track the evolution of energy densities of
1. radiation, 2. LSPs $\tchi$, 3. CO-produced axions, 4.
thermally- and decay-produced axions, 5. CO-produced saxions, 6. thermally- and decay-produced saxions, 7. axinos and 8. gravitinos in the early universe from
the epoch of reheat to the era of entropy conservation\cite{Bae:2014rfa}.
Applied to (the more plausible) case of natural SUSY models, as typified
by low finetuning measure $\Delta_{EW}$\cite{Baer:2012up,Baer:2012cf} and with higgsino-like LSPs,
it is typically found that the dark matter is axion-dominated with just
a small component-- usually 10-20\%-- of neutralino dark matter\cite{Bae:2014rfa}.
The reduced neutralino abundance allows the natural higgsino-like
relic WIMPs to escape stringent bounds from direct and indirect detection
experiments\cite{Baer:2016ucr}. In addition, the $a\gamma\gamma$ coupling in the MSSM
is severely reduced\cite{Bae:2017hlp} allowing SUSY DFSZ axions to lie well hidden below
the present reach of axion haloscope searches\cite{ADMX:2021nhd}.

To address problem \#3 above, unifying the SM with gravity, it seems
necessary to embed the SUSY DFSZ setup into the string theory framework\cite{Ibanez:2012zz,Hebecker:2021egx}.
In string theory, under flux compactifications\cite{Douglas:2006es}, a vast number of vacuum configurations emerge, each giving rise to distinct $4-d$ laws of physics.
The number of distinct vacua has been estimated as $N_{vac}\sim 10^{500}$\cite{Ashok:2003gk},
but other estimates can produce many more\cite{Taylor:2015xtz}.
The vast landscape\cite{Susskind:2003kw} of string vacua can all be accessed
in the model of an eternally inflating
multiverse\cite{Guth:2000ka,Linde:2015edk}, which provides a setting for
Weinberg's anthropic solution to the cosmological constant problem\cite{Weinberg:1987dv,Bousso:2000xa},
thus explaining the tiny magnitude of measured dark energy.
The string landscape picture, applied to supersymmetric models, also seems
to favor SUSY models with large soft terms\cite{Douglas:2004qg,Susskind:2004uv,Arkani-Hamed:2005zuc} while respecting low $\Delta_{EW}$
via the requirement of the pocket-universe value of the weak scale lying
within the Agrawal {\it et al.}\cite{Agrawal:1998xa} allowed window of
values which lead to complex nuclei as in our universe (atomic principle).
Thus, the string landscape should statistically favor natural SUSY models
which generate $m_h\sim 125$ GeV with sparticles beyond present LHC
bounds\cite{Baer:2016lpj,Baer:2017uvn,Baer:2022naw}.

A consequence of string theory compactifications on Calabi-Yau threefolds or M-theory compactifications on manifolds with $G_2$ holonomy
is that, in addition to MSSM fields (plus other possible exotica),
one expects the generic presence of moduli fields:
gravitationally coupled scalar fields (string remnants)
which parameterize the size and shape of the extra string dimensions within
the 6-7 dimensional compact space. The moduli can be categorized as the
dilaton $S$ plus Hodge number $h^{2,1}$ complex structure (shape) moduli $U^\beta$ and
$h^{1,1}$ K\"ahler (size) moduli $T^\alpha$.
Realistic string models require stabilized moduli since their vevs determine
many quantities in the $4-d$ low energy effective field theory (LE-EFT) such
as gauge and Yukawa couplings and soft SUSY breaking terms.
Additionally, unstabilized moduli appear as massless scalar fields in the $4-d$ theory, leading to (unobserved) long-range fifth forces.
In the Type IIB string context, the $S$ and $U^\beta$ can be stabilized via flux\cite{Giddings:2001yu} whilst
the $T^\alpha$ may be stabilized by non-perturbative effects\cite{Kachru:2003aw} (gaugino condensation, string instantons) in the KKLT scenario
or via a combination of perturbative and non-perturbative effects
in the large volume scenario (LVS) \cite{Balasubramanian:2005zx}.
While the $S$ and $U^\beta$ moduli gain masses of order the Kaluza-Klein scale
(at least in the IIB context), 
the K\"ahler moduli may be much lighter, of order the SUSY breaking
scale\cite{Acharya:2010af}, hierarchically heavier than the SUSY breaking scale by a
factor $[\log (m_P/m_{3/2})]^2$\cite{Choi:2005ge}, or several orders of magnitude heavier than the SUSY breaking scale - but still far below the Kaluza-Klein scale - in sequestered scenarios \cite{Blumenhagen:2009gk, Aparicio:2014wxa, Cicoli:2015bpq}.

The presence of light moduli, as seems generic in string
compactifications\cite{Acharya:2010af},
may give rise to the cosmological moduli problem
(CMP)\cite{Coughlan:1983ci,Banks:1993en,deCarlos:1993wie,Kane:2015jia}
wherein the lightest
modulus field $\phi$ can be produced via CO at a temperature $T_{osc}$
and thereafter may quickly come to dominate the energy density of the universe since their energy density scales as $R^{-3}$ while the energy density of
radiation scales as $R^{-4}$. Since $\phi$ is gravitationally coupled,
it is long-lived and could potentially decay after BBN, thus destroying the
successful prediction of light element abundances.
Also, $\phi$ can decay to gravitinos which in turn could decay after BBN,
or else overproduce LSP dark matter\cite{Hashimoto:1998mu,Kohri:2004qu,Nakamura:2006uc,Asaka:2006bv,Endo:2006zj},
or they could directly decay to SUSY particles,
again overproducing LSP dark matter\cite{Blinov:2014nla}.
If $\phi$ is coupled to light axion-like particles (ALPs), then they
potentially can overproduce dark radiation (DR)\cite{Cicoli:2012aq,Higaki:2012ar,Cicoli:2015bpq,Baer:2022fou}.

In previous work, we have computed all modulus decay modes to MSSM particles
including all mixing and phase space effects, and assessed solutions to the
CMP\cite{Bae:2022okh}.
A very plausible solution is that $\phi$ is so heavy that it decays before
the onset of BBN, or even before WIMP freeze-out.
This modulus decoupling solution seems to require $m_\phi\agt 2.5\times 10^3$ TeV.
To avoid the moduli-induced gravitino problem, one possibility is to require $m_\phi<2m_{3/2}$.
Another possibility, as discussed by Dine \textit{et al.} \cite{Dine:2006ii}, 
is that if the light modulus has a large supersymmetric mass and direct couplings with any SUSY-breaking hidden sector fields are absent, 
the offending unsuppressed decays of the modulus to gravitinos cancel, leaving only the helicity-suppressed contribution.
In this case, heavy moduli decay to gravitinos with a very small branching ratio $\ll 1\%$.
In cases where the modulus mass $m_\phi$ is linked to the SUSY breaking scale,
then one would expect SUSY breaking of order $10^3$ TeV, giving rise to
huge values of $\Delta_{EW}$\cite{Baer:2012up} and an obvious conflict with
electroweak naturalness.
This problem is naturally absent in scenarios that exhibit sequestering \cite{Randall:1998uk,DeWolfe:2002nn,Berg:2010ha,Kachru:2007xp,Aparicio:2014wxa,Blumenhagen:2009gk}.
If the MSSM is confined to $D3$-branes and SUSY breaking originates elsewhere,
this effect must be communicated across the bulk - allowing a gravitino with
extremely large mass while soft terms can remain in the TeV range.
This scenario also solves the moduli-induced gravitino problem in LVS
scenarios \cite{Cicoli:2012aq}, as the bulk modulus is lighter than the
gravitino and so this problematic decay is kinematically inaccessible.
Alternatively, an anthropic solution to the CMP was suggested in
Ref. \cite{Baer:2021zbj} which requires the modulus field amplitude $\phi_0$
in our pocket universe to be anthropically selected to be
$\phi_0\alt 10^{-7}m_P$ in order
to avoid a pocket universe with too large a dark matter to baryonic
matter ratio which would lead to structure being dominated by DM instead of
baryons\cite{Wilczek:2004cr,Tegmark:2005dy,Hertzberg:2008wr,Freivogel:2008qc}.
In Ref. \cite{Baer:2022fou}, we examined
the issue of dark radiation when the axionic component of $\phi$ is a light ALP,
as expected in LVS.

In the present paper, we move a step further and examine the (well-motivated)
case of a light modulus coupled  to the PQMSSM (Peccei-Quinn augmented MSSM),
so that MSSM particles, a light modulus $\phi$, and axions, axinos, saxions
and gravitinos are all present in the early universe.
For the PQ sector, an important issue is the origin of the required
global $U(1)_{PQ}$ symmetry since global symmetries are not compatible with
string theory\cite{Banks:1988yz}.
A perhaps related problem is the so-called
{\it axion quality problem}\cite{Kamionkowski:1992mf,Barr:1992qq} where
non-renormalizable and/or non-perturbative contributions to the axion potential
can displace its minimum enough so that the bound $\bar{\theta}\alt 10^{-10}$
is no longer respected.
A further problem for the case of stringy axions (specifically, closed string axions)
is that the magnitude of the decay constant $f_a$ (usually) turns out too high--
$f_a\sim 10^{16}$ GeV-- as it is linked to the string scale\cite{Svrcek:2006yi}.
Such a high value of $f_a$ leads to overproduction of CO-produced axions
and in SUSY PQ models, it leads to overproduction of WIMP DM and possibly
violation of BBN and DR bounds\cite{Bae:2014rfa}.
One solution in IIB string theory is to consider instead open string axions such as the construction in Ref. \cite{Cicoli:2013cha}, which takes the QCD axion to be the phase of some PQ matter field residing on a $D3$-brane, instead of arising from dimensional reduction of the Ramond-Ramond sector gauge fields (see {\it e.g.} Ref. \cite{Cicoli:2012sz}).
By considering open string axions, the vev of the PQ field - and thus the decay constant $f_a$ - can arise at a much lower scale than their closed string counterparts.
Open string constructions of the SUSY DFSZ model arising from discrete symmetries have also been considered in the Type IIA context in Refs. \cite{Honecker:2013mya,Honecker:2015ela}.
Additionally, a similar scenario was studied in Ref. \cite{Cicoli:2022uqa} where, in the context of the fibred LVS framework with parameters chosen from cosmological data, a viable inflationary scenario with a light post-inflationary modulus was shown to predict higgsino-like WIMPs, TeV-scale soft terms, and potentially non-negligible contributions to dark radiation from open-string axions.

We adopt the approach advocated in Ref. \cite{Choi:2009jt,Nilles:2017heg} where the breaking of
higher dimensional Lorentz symmetry in the string compactification can lead to
remnant discrete $R$-symmetries, which are compatible with string theory.
$R$-symmetries are intrinsically supersymmetric since the anti-commuting
superspace dimensions transform non-trivially, and lend further credence
to how SUSY helps solve a variety of issues present in non-SUSY PQ models.
In Ref. \cite{Lee:2011dya}, all anomaly-free discrete $R$-symmetries compatible with
(local\cite{Buchmuller:2005sh}) grand unification were tabulated.
In Ref. \cite{Baer:2018avn},
two SUSY DFSZ axion models based on a ${\bf Z}_{24}^R$ discrete symmetry
were presented (see also Ref's. \cite{Bhattiprolu:2021rrj} and \cite{Choi:2022fha}).
In these models, the $U(1)_{PQ}$
emerges as an accidental, approximate global symmetry as a consequence
of the more fundamental discrete $R$-symmetry. The ${\bf Z}_{24}^R$
symmetry is sharp enough to suppress higher dimensional operators
up to $m_P^{-7}$ in the superpotential,
thus solving the axion quality problem\footnote{The presence of string instantons may also disrupt the axion quality; for recent discussion, see {\it e.g.}
Refs. \cite{Demirtas:2021gsq,Dine:2022mjw}.}.
Also, in this case the $U(1)_{PQ}$ is spontaneously broken as a consequence of
SUSY breaking; this leads to $f_a \sim \sqrt{m_{soft} m_P}\sim 10^{11}$ GeV,
in the cosmological sweet zone for axion production (where $m_{soft}$ is the
scale of MSSM SUSY breaking, $\sim$ 1-10 TeV).
In this context, the axion emerges as an accion, as in Ref. \cite{Choi:2009jt}.

The remainder of this paper is organized as follows. In Sec. \ref{sec:model1},
we write down our model for the modulus coupling to the PQ sector and
extract the associated Lagrangian. This is used to compute modulus decay widths
to axions, saxions and axinos. The connection of our coupling to the
gravity-safe ${\bf Z}_{24}^R$ model is detailed in Sec. \ref{ssec:model2}.
In Sec. \ref{sec:BFs}, we present numerical results for modulus decay
to PQMSSM particles. For comparable couplings, the modulus dominantly decays
via $\phi\to aa$ and $ss$ whilst decay to $\ta\ta$ is generically helicity suppressed.
For the aid to the reader, we also show associated branching fractions for
saxions and gravitinos.
In Sec. \ref{sec:Boltz}, we extend our previous calculations of dark matter production
within the PQMSSM model to include a ninth Boltzmann equation for the light
modulus field. We show plots of how the early universe constituent energy
densities evolve with increasing scale factor from the time of reheat
to the era of entropy conservation.
In Sec. \ref{sec:Oh2_phiPQ}, we scan over $\phi$PQMSSM parameter space to locate
regions where viable amounts of dark matter are produced, where there is
not too much DR and where the CMP is solved.
A summary and conclusions are presented in Sec. \ref{sec:conclude}.

\section{Modulus coupled to the PQ sector}
\label{sec:model1}

We consider first the modulus coupling to the PQ-charged superfield,
$\hat{\sigma}$, within the SUSY DFSZ framework.
The leading allowed interaction is from the K\"ahler potential:
\begin{equation}
    \mathcal{L}
    \supset
    \int
    d^4\theta
    \, 
    \left[
        \frac{
            \lambda_{PQ}
        }{
            m_P
        }
        \hat{\Phi}
        \hat{\sigma}^\dagger
        \hat{\sigma}
        +
        \text{h.c.}
    \right]
\end{equation}
where $\hat{\Phi}$ is the modulus superfield.
The PQ symmetry is then postulated to break at the scale $\sim f_a$ and
$\sigma$ acquires a VEV, at which point fluctuations in the phase take on the role
of the axion superfield.
Integrating out the heavy PQ field, this interaction can then be parameterized
as
\begin{equation}
    \mathcal{L}
    \supset
    v_{PQ}^2
    \frac{
        \lambda_{PQ}
    }{
        m_P
    }
    \int
    d^4\theta
    \,
    \left[
        \hat{\Phi}
        \exp 
        \left(
            \frac{q}{f_a}
            \left[
                \hat{A}
                +
                \hat{A}^\dagger
            \right]
        \right)
        +
        \text{h.c.}
    \right]
\end{equation}
where $q$ is the PQ charge of $\hat{\sigma}$, $v_{PQ} = \langle \sigma \rangle$, and $\hat{A}$ is the axion superfield.
The modulus decay terms are then given by expanding the above form\footnote{
    Expanding this form to first order leads to kinetic mixing effects between $\hat{\Phi}$ and $\hat{A}$.
    These effects take the form $\lambda_{PQ} q v_{PQ}^2 / ( m_P f_a ) \hat{\Phi} \hat{A}^\dagger + \text{h.c.}$ and may be removed with a suitable field redefinition.
    We ignore this effect in this work, as it leads to a correction in the kinetic terms $\Phi^\dagger \Phi \rightarrow (1 + f_a^2/m_P^2) \Phi^\dagger \Phi \sim \Phi^\dagger \Phi$.
}:
\begin{equation}
    \mathcal{L}
    \supset
    \frac{1}{2}
    \frac{
        \lambda_{PQ}
    }{
        m_P
    }
    \int 
    d^4 \theta
    \,
    \left(
        \hat{\Phi}
        +
        \hat{\Phi}^\dagger
    \right)
    \left(
        \hat{A}
        +
        \hat{A}^\dagger
    \right)^2
\end{equation}
where we have used $f_a^2 = q^2 v_{PQ}^2$.
(We may just as well have performed a field redefinition of $\hat{A}$ to
obtain canonical kinetic terms arising from a similar parameterization
of the PQ field's kinetic term, $K \supset \hat{\sigma}^\dagger \hat{\sigma}$).
Evaluating the superspace integral and expanding superfields into components leads to the Lagrangian
\begin{multline}
    \label{eq:fullmodulusPQLagrangian}
    \mathcal{L}_{\phi A A}
    \supset
    \frac{1}{2}
    \frac{
        \lambda_{PQ}
    }{
        m_P
    }
    \left[
        -
        2
        \Phi 
        A
        \partial^2
        A^\dagger
        -
        A^\dagger
        A^\dagger
        \partial^2
        \Phi
    \right]
    +
    i
    \frac{
        \lambda_{PQ}
    }{
        m_P
    }
    \left[
        \Phi 
        \overline{
            \tilde{a}
        }
        \slashed{\partial}
        P_R
        \tilde{a}
    \right]
    \\
    +
    \frac{
        \lambda_{PQ}
    }{
        m_P
    }
    \left[
        \Phi 
        F_A
        F_A^\dagger
        +
        (
            A
            +
            A^\dagger
        )
        F_\phi 
        F_A^\dagger
        +
        \frac{i}{2}
        F_\phi
        \overline{\tilde{a}}
        P_R 
        \tilde{a}
    \right]    
    +
    \text{h.c.}
\end{multline}
where the $F_i$ are auxiliary fields and where
we ignore the modulino, $\psi_\phi$.
Expanding into the saxion and axion components, $\Phi = ( \phi + i c ) / \sqrt{2}$ and $A = ( a + i s ) / \sqrt{2}$, and focusing on the modulus interactions,
we recover the Lagrangian 
\begin{equation}
    \mathcal{L}_{\phi A A}
    \supset
    -
    \frac{
        \lambda_{PQ}
    }{
        2
        \sqrt{2}
        m_P
    }
    \left[
        s
        s
        \partial^2
        \phi 
        +
        2 
        \phi 
        s
        \partial^2
        s
        -
        a
        a
        \partial^2
        \phi 
        +
        2
        \phi
        a
        \partial^2
        a
    \right]
    +
    i
    \frac{
        \lambda_{PQ}
    }{
        \sqrt{2}
    }
    \phi 
    \overline{\tilde{a}}
    \slashed{\partial}
    \tilde{a}
    .
\end{equation}
Note that, integrating by parts the $\phi a a$ couplings leads to interactions
of the form $\phi \, \partial_\mu a \, \partial^\mu a$, and hence the required axionic shift
symmetry is indeed present.
The above Lagrangian can be used to compute the following decay widths: 
\begin{align}
    \Gamma
    \left(
        \phi 
        \rightarrow
        s
        s
    \right)
    &=
    \frac{
        \lambda_{PQ}^2
    }{
        64\pi
    }
    \frac{
        m_\phi^3
    }{
        m_P^2
    }
    \left(
        1
        +
        2
        \frac{
            m_s^2
        }{
            m_\phi^2
        }
    \right)^2
    \lambda^{1/2}
    \left(
        1,
        \frac{
            m_s^2
        }{
            m_\phi^2
        },
        \frac{
            m_s^2
        }{
            m_\phi^2
        }
    \right)
    \\
    \Gamma
    \left(
        \phi 
        \rightarrow
        a
        a
    \right)
    &=
    \frac{
        \lambda_{PQ}^2
    }{
        64\pi
    }
    \frac{
        m_\phi^3
    }{
        m_P^2
    }
    \left(
        1
        -
        2
        \frac{
            m_a^2
        }{
            m_\phi^2
        }
    \right)^2
    \lambda^{1/2}
    \left(
        1,
        \frac{
            m_a^2
        }{
            m_\phi^2
        },
        \frac{
            m_a^2
        }{
            m_\phi^2
        }
    \right)
    \\
    \Gamma
    \left(
        \phi 
        \rightarrow
        \overline{\tilde{a}}
        \tilde{a}
    \right)
    &
    =
    \frac{
        \lambda_{PQ}^2
    }{
        8 \pi 
    }
    \frac{
        m_\phi^3
    }{
        m_P^2
    }
    \left(
        \frac{
            m_{\tilde{a}}^2
        }{
            m_\phi^2
        }
    \right)
    \left(
        1
        -
        4
        \frac{
            m_{\tilde{a}}^2
        }{
            m_\phi^2
        }
    \right)
    \lambda^{1/2}
    \left(
        1,
        \frac{
            m_{\tilde{a}}^2
        }{
            m_\phi^2
        },
        \frac{
            m_{\tilde{a}}^2
        }{
            m_\phi^2
        }
    \right)
    .
\end{align}
Here, the $\phi\to \ta\ta$ decay receives helicity suppression.
However, the decay widths to saxion and axion pairs are unsuppressed,
and are thus the leading $\phi$ decay modes to the PQ sector.
The remaining modulus decay widths to the various MSSM particles
can be found in the Appendix to Ref. \cite{Bae:2022okh}.

Before we proceed, we note that the $F$-term interactions with the axinos
can lead to (model-dependent) unsuppressed axino widths.
Namely, if the modulus is stabilized supersymmetrically, as in KKLT, upon integrating out all the heavy fields one could parameterize the modulus mass with the superpotential term $W \supset \int d^2 \theta \, M_\Phi \Phi^2$ where $M_\Phi$ is the supersymmetric mass of $\Phi$.
The $F$-term interaction in Eq.~(\ref{eq:fullmodulusPQLagrangian}) then leads to the interaction $(i \lambda_{PQ} / 2 m_P) M_\Phi \Phi \overline{\tilde{a}} P_R \tilde{a} + \text{h.c.} $, which is an unsuppressed contribution.
However, in {\it e.g.} LVS where supersymmetry is broken by the lightest modulus, the modulus mass is not supersymmetric so that $M_\Phi \sim 0$ and we are left with only the suppressed interaction.
This type of $F$-term interaction also leads to unsuppressed decays to gauginos through the gaugino mass term, which we referred to as cases \textbf{A1} and \textbf{A2} in Ref. \cite{Bae:2022okh} (where \textbf{A1} and \textbf{A2} have unsuppressed and suppressed decays to gravitinos respectively, which are also model-dependent decays originating from the same details of the moduli-hidden sector interactions).
We focus solely on the suppressed axino+gaugino case in this work, and leave detailed treatment of the unsuppressed case for future work. 

\subsection{Connection to gravity-safe PQ models (GSPQ)}
\label{ssec:model2}

Due to the absence of global symmetries in string theory, one may take issue with the appearance of a global $U(1)_{PQ}$ symmetry at this scale and question the viability of the above model.
Here we connect the above model to the class of PQ models based on discrete $R$-symmetries - which \textit{are} expected to be compatible with string theory.
The GSPQ models introduce PQ-charged superfields $\hat{X}$ and $\hat{Y}$, and introduce charge assignments which dictate the field content respects an approximate $U(1)_{PQ}$ symmetry in the superpotential at $\mathcal{O}(m_P^{-1})$, while the next leading operators are suppressed by at least $\mathcal{O}(m_P^{-7})$ - which suppresses contributions from PQ breaking terms in the scalar potential by at least $\mathcal{O}(m_P^{-8})$.
This large suppression was found to be sufficient for gravity safety in Ref. \cite{Kamionkowski:1992mf}.


In our approach, we note that for each of the GSPQ models studied in Ref. \cite{Baer:2018avn}, the following operators are allowed in the K\"{a}hler potential:
\begin{equation}
    \mathcal{L}_{PQ}
    \supset
    \int 
    d^4
    \theta 
    \, 
    \left[
        \frac{
            \lambda_X
        }{
            m_P
        }
        \hat{\Phi} 
        \hat{X}^\dagger
        \hat{X}
        +
        \frac{
            \lambda_Y
        }{
            m_P
        }
        \hat{\Phi} 
        \hat{Y}^\dagger
        \hat{Y}
        +
        \text{h.c.}
    \right]
\end{equation}
where we assume that the modulus is uncharged under the discrete $R$-symmetry (and hence uncharged under the approximate $U(1)_{PQ}$ symmetry).
Once the $X$ and $Y$ fields acquire their VEVs and the approximate PQ symmetry has been broken, we may integrate out the $X$ and $Y$ fields which leads to the form 
\begin{equation}
    \mathcal{L}_{PQ}
    \supset
    \int 
    d^4
    \theta 
    \, 
    \left[
        \frac{
            \lambda_X
            v_X^2
        }{
            m_P
        }
        \hat{\Phi} 
        \exp 
        \left(
            \frac{
                q_X
            }{
                f_a
            }
            \left(
                \hat{A}
                +
                \hat{A}^\dagger
            \right)
        \right)
        +
        \frac{
            \lambda_Y
            v_Y^2
        }{
            m_P
        }
        \hat{\Phi} 
        \exp 
        \left(
            \frac{
                q_Y
            }{
                f_a
            }
            \left(
                \hat{A}
                +
                \hat{A}^\dagger
            \right)
        \right)
        +
        \text{h.c.}
    \right] .
\end{equation}
Expanding the exponential to second-order leads to the interactions
\begin{equation}
    \mathcal{L}_{PQ}
    \supset
    \frac{
        1
    }{
        2 m_P
    }
    \left[
        \frac{
            \lambda_X
            q_X^2
            v_X^2
            +
            \lambda_Y
            q_Y^2
            v_Y^2
        }{
            f_a^2
        }
    \right]
    \int 
    d^4
    \theta 
    \, 
    \left(
        \hat{\Phi} 
        +
        \hat{\Phi}^\dagger
    \right)
    \left(
        \hat{A}
        +
        \hat{A}^\dagger
    \right)^2
    .
\end{equation}
This is identical to the form of our simple model in the previous section,\footnote{It is straightforward to show that the kinetic terms, $\hat{X}^\dagger \hat{X}$ and $\hat{Y}^\dagger \hat{Y}$, lead to a canonically normalized kinetic term for the axion superfield, $\hat{A}^\dagger \hat{A}$, since $f_a^2 = \sum_i q_i^2 v_i^2$.}
with an effective coupling 
\begin{equation}
    \lambda_{PQ}
    =
    \frac{
        \lambda_X
        q_X^2
        v_X^2
        +
        \lambda_Y
        q_Y^2
        v_Y^2
    }{
        f_a^2
    }
\end{equation}
which, assuming $\mathcal{O}(\lambda_X) \sim \mathcal{O}(\lambda_Y) \sim \mathcal{O}(1)$, we see that $\lambda_{PQ} \sim \mathcal{O}(1)$.
The GSPQ models of Ref. \cite{Baer:2018avn} thus reduce to our simple model once PQ
symmetry is broken, suggesting that our simple model is sufficient to describe
a wide class of PQ models believed to be compatible with string theory.

\section{Modulus, saxion and gravitino branching fractions}
\label{sec:BFs}

\subsection{A natural  MSSM benchmark model}
\label{ssc:BM}

To illustrate the modulus branching fractions to PQMSSM particles, we adopt
the same natural SUSY benchmark model (BM) as in our previous works,
\cite{Bae:2022okh}. This BM point is taken from the three-extra-parameter
non-universal Higgs model NUHM3\cite{Baer:2005bu},
with parameters $m_0(1,2)=10$ TeV, $m_0(3)=5$ TeV,
$m_{1/2}=1.2$ TeV, $A_0=-8$ TeV, $\tan\beta =10$ with $\mu =200$ GeV
and $m_A=2$ TeV. We use the computer code Isajet 7.88\cite{Paige:2003mg} to
generate the spectra. A full Table of the Higgs and sparticle mass
spectra is shown in Ref. \cite{Bae:2022okh} and so we do not repeat it here.
We do remark that $m_{\tg}\simeq 2.9$ TeV and $m_{\tst_1}\simeq 1.25$ TeV
with $m_h=125.3$ GeV so the BM model is consistent with current LHC constraints.
The lightest electroweakinos are higgsino-like with mass $\sim 200$ GeV
and the electroweak naturalness measure\cite{Baer:2012up} $\Delta_{EW}=20$
so the model is natural.
The thermally-produced (TP)  LSP relic abundance is
$\Omega_{\tchi}^{TP}\sim 0.011$ (using IsaReD\cite{Baer:2002fv}) \footnote{
    The thermally-produced value we find from numerical solution of the Boltzmann equations is $\Omega_{\tchi} h^2 = 0.0044$.
    The discrepancy with the IsaReD output is due to the fact that the IsaReD routine uses only a temperature-independent value of the cross section close to freeze-out for its estimate, in addition to a semi-analytic formula.
    Close to freeze-out, our benchmark point still has appreciable temperature-dependent contributions to the annihilation cross section which are used in our Boltzmann code.
    We expect our Boltzmann code to give a more accurate thermal-value in this case, as it closely tracks the equilibrium density until freeze-out.
}
and is thus underabundant;
however, the TP LSP relic density will be drastically changed under the
presence of both a light modulus and the PQ sector. 
Throughout this work, we adopt the case {\bf B2} from Ref. \cite{Bae:2022okh} where modulus decays
to both gaugino pairs and gravitinos pairs are helicity suppressed.

We also stipulate several PQMSSM parameters which enter into the ensuing plots. 
For our PQMSSM benchmark model, we take $f_a=10^{11}$ GeV with $m_s=m_{\ta}=5$ TeV, 
and initial saxion amplitude as $s_i=f_a$ with $\theta_i=3.113$ and PQ effective self-coupling
$\xi =1$. 

Finally, we consider string-inspired expectations for the magnitude of the modulus coupling to the gauge sector, $\lambda_{gauge}$, which depends on the details of the gauge-kinetic function.
For the case where the lightest modulus appears in the gauge-kinetic function at tree level, we take $\lambda_{gauge}=1$ and refer to this as case \textbf{GK1}.
This case might arise from compactifications of the heterotic string wherein the gauge-kinetic function depends primarily on the dilaton.
In heterotic cases, fluxes may not be able to stabilize the dilaton (due to the absence of the Ramond-Ramond fields, which are present in the IIB setting) and so the lightest modulus may be the dilaton in this context \cite{Lowen:2008fm,Lowen:2008xh}.
This case might also arise in M-theoretic compactifications on $G_2$ manifolds, where the gauge-kinetic function is set by all geometric moduli \cite{Acharya:2007rc}.
Additionally, this case could arise within the context of the Type IIB setting if the MSSM resides on $D7$ branes in the geometric regime, although this particular scenario might require extreme tuning of the flux superpotential $W_0$ and could possibly suffer from the CMP \cite{Conlon:2006wz,AbdusSalam:2007pm,Blumenhagen:2009gk}.
The second case which we refer to as \textbf{GK2} is motivated from scenarios where the lightest modulus has a suppressed contribution in the gauge kinetic function.
Here, the notable example is Type IIB string compactifications where the MSSM resides on $D3$ branes located at singularities.
In this case, the gauge-kinetic function is again set predominantly by the dilaton, while dependence of the gauge-kinetic function on the lightest modulus might then appear at loop-level \cite{Blumenhagen:2009gk,Cicoli:2012aq,Aparicio:2014wxa}.
Although from different arguments, this scenario is also mentioned by Moroi and Randall \cite{Moroi:1999zb} since, unless the lightest modulus has a vanishing or highly suppressed $F$-term, the gaugino mass will then be pushed to large values.
For this case, we take a loop-suppressed gauge coupling $\lambda_{gauge}=1/16\pi^2$.
Both of these cases are listed in Table \ref{tab:casesGaugeKinetic} for the aid of the reader.

\begin{table}[ht!]
    \centering
    \begin{tabular}{c | c | c |}
    & Tree-level/unsuppressed & Loop-level/suppressed \\
    \hline
    Gauge-kinetic function & Case \textbf{GK1}, $\lambda_{gauge} = 1$ & Case \textbf{GK2}, $\lambda_{gauge} = 1/16\pi^2$ \\
    \hline
    \end{tabular}
    \caption{Summary of case scenarios for the expected magnitude of $\lambda_{gauge}$.
    }
    \label{tab:casesGaugeKinetic}
\end{table}

\subsection{Modulus branching fractions}
\label{ssec:mBFs}

\begin{figure}[tbh!]
    \centering
    \includegraphics[scale=0.48]{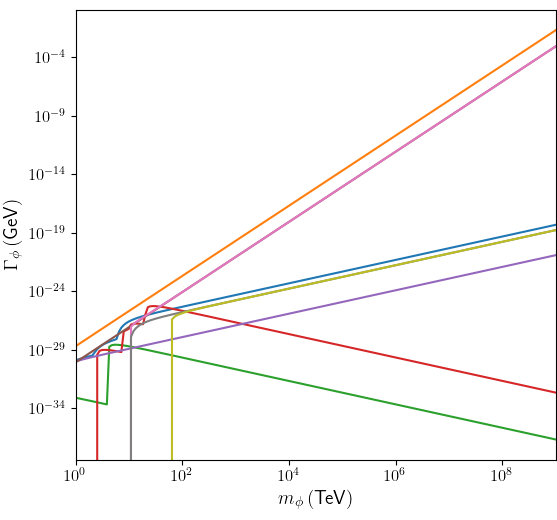}
    \includegraphics[scale=0.48]{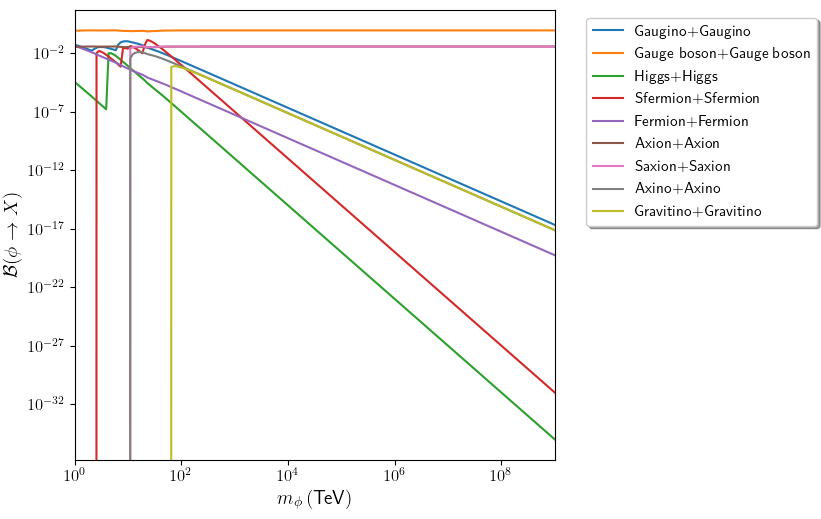}
    \caption{Lightest modulus field $\phi$ {\it a}) decay widths and
    {\it b}) branching fractions
    into PQMSSM particles  vs. $m_\phi$ for case \textbf{GK1}.
    The MSSM particles have masses as in the above described
    natural SUSY BM point while we take the gravitino mass $m_{3/2}=30$ TeV.
    For PQ sector particles, we take $m_{\ta}=m_s=5$ TeV.
    We take all PQ couplings $\lambda_{PQ}=1$, as well as $\lambda_{gauge} = 1$
    and all MSSM couplings
    $\lambda_i =1$ (where $i$ runs over the various modulus-MSSM couplings\cite{Bae:2022okh}) and with the case of helicity-suppressed $\phi$ decays to
    gauginos and gravitinos.
    \label{fig:phiBF1}}
\end{figure}

We use the above modulus decay widths along with their MSSM counterparts
to compute the light modulus field $\phi$ decay widths into
PQMSSM particles for the case \textbf{GK1} in Fig. \ref{fig:phiBF1}{\it a}).
The MSSM particles have masses as in the above described
natural SUSY BM point while
we take the gravitino mass $m_{3/2}=30$ TeV. For PQ sector particles,
we take the axino mass $m_{\ta}=5$ TeV and the saxion mass $m_s=5$ TeV.
We take all PQ couplings $\lambda_{PQ}=1$ and all MSSM couplings
$\lambda_i =1$ (where $i$ runs over the various modulus-MSSM couplings\cite{Bae:2022okh}). PQ symmetry forbids the existence of the Giudice-Masiero (GM)
term, so for the entirety of this work, we set $\lambda_H = 0$,
and exclude this coupling from consideration when we made adjustments
to all other couplings $\lambda_i$.

From the plot, we see that the dominant $\Gamma_\phi$ decay width is into
vector boson pairs $VV$: $WW$, $ZZ$, $\gamma\gamma$, $\gamma Z$ and $gg$.
The second largest decay widths are into saxion pairs $\phi\to ss$ and axion pairs $\phi \to aa$ (which overlap in the figure, except for $m_\phi \lesssim 2 m_s = 10$ TeV). 
In frame {\it b}), we show the corresponding modulus branching fractions.
The dominant $\phi$ branching fraction is into vector boson pairs, $VV$,
which can lead to large entropy dilution of all relics in the early universe
present at the time of modulus field decay. The second largest branching
fraction is into saxion (and axion) pairs. The saxion decay mode can lead to
1. entropy dilution at the time of saxion decay, 2. additional WIMP
production from saxion decay and 3. dark radiation production if
$s\to aa$ occurs.
Additionally, the decay into axions leads to production of dark radiation.

\begin{figure}[tbh!]
    \centering
    \includegraphics[scale=0.48]{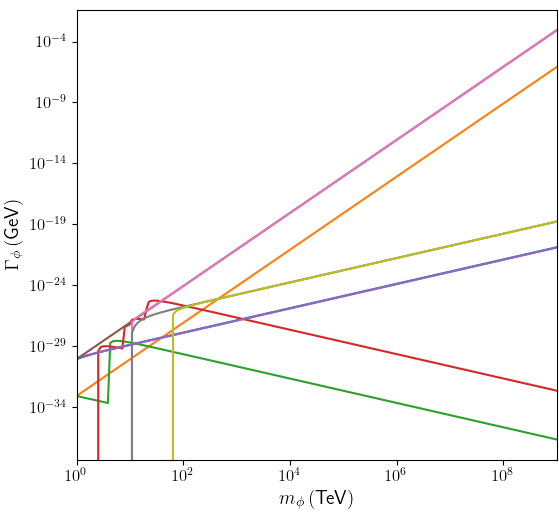}
    \includegraphics[scale=0.48]{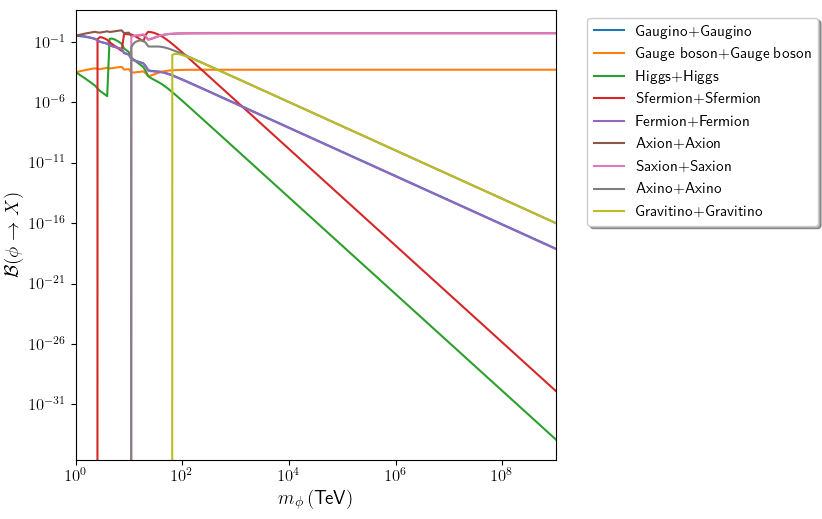}
    \caption{
        Lightest modulus field $\phi$ {\it a}) decay widths and
        {\it b}) branching fractions
        into PQMSSM particles  vs. $m_\phi$ for case \textbf{GK2}. This case takes $\lambda_{\rm PQ}=1$
        and all other moduli
        couplings $\lambda_i =1$ except $\lambda_{gauge}=1/16\pi^2$. 
        We adopt again the case of helicity-suppressed decays to
        gauginos and gravitinos for a natural SUSY BM point from
        Ref. \cite{Bae:2022okh}.
    \label{fig:phiBF2}}
\end{figure}

In Fig. \ref{fig:phiBF2}, we show the ensuing modulus decay widths and
branching fractions for the case \textbf{GK2}.
With $\lambda_{gauge}$ suppressed, now $\phi\to ss$ and $\phi \to aa$ are the dominant modulus
decay modes. (These two decay modes again overlap in the figure except where $\phi \rightarrow s s$ is kinematically forbidden).
In this case, dark matter production in the early universe
will depend heavily on the saxion decay modes.

\subsection{Saxion branching fractions}
\label{ssec:sBFs}

As shown in Refs. \cite{Bae:2013hma,Chun:1995hc},
the axion-axino-saxion kinetic terms and self-couplings
(in four component notation) are of the form
\be
{\cal L}=\left(1+\frac{\sqrt{2}\xi}{v_{PQ}}s\right)
\left[\frac{1}{2}\partial^\mu a\partial_\mu a+\frac{1}{2}\partial^\mu s\partial_\mu s
+\frac{i}{2}\bar{\ta}\dsl \ta\right]
\ee
where $\xi =\sum_i q_i^3v_i^2/v_{PQ}^2$.
Here $q_i$ and $v_i$ denote PQ charges and vacuum expectation values of PQ
fields $S_i$, and the PQ scale $v_{PQ} = f_a/\sqrt{2}$ 
is given by $v_{PQ}=\sqrt{\sum_i q_i^2v_i^2}$.
In the above interaction, $\xi$ is typically $\sim 1$, but in some cases can
be as small as $\sim 0$~\cite{Chun:1995hc}.

\begin{figure}[tbh!]    
    \centering
    \includegraphics[scale=0.48]{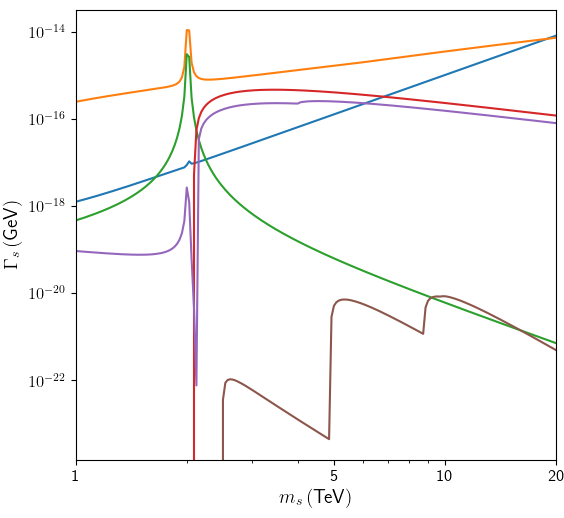}
    \includegraphics[scale=0.48]{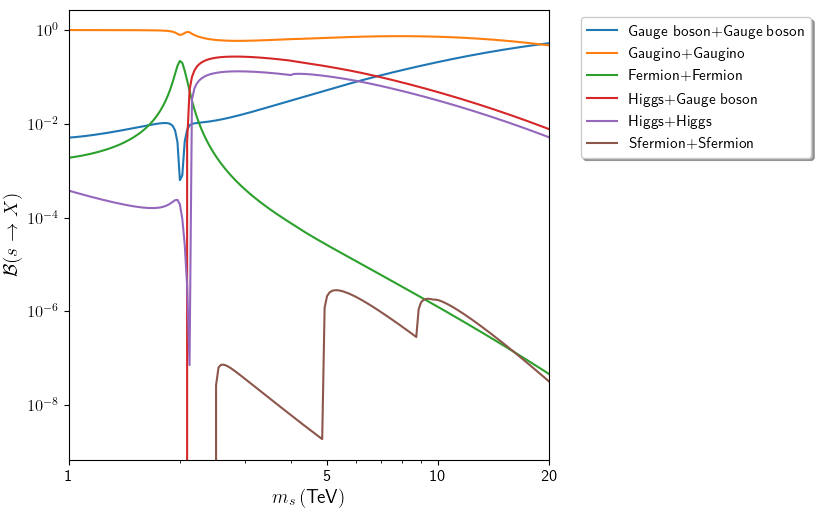}
    \caption{Saxion $s$  {\it a}) decay widths and
    {\it b}) branching fractions into PQMSSM particles  vs. $m_s$ for $\xi=0$.
    \label{fig:sBFxi0}}
\end{figure}

The saxion decay widths and branching fractions are shown in
Fig. \ref{fig:sBFxi0} vs. saxion mass $m_s$ for the natural SUSY BM with
$m_{\ta}=5$ TeV and taking $\xi =0$ (decay to PQ sector suppressed).
The corresponding formulae for saxion decay widths are listed in
Ref. \cite{Bae:2013hma}. From the plots, we see for $\xi=0$ that the dominant
saxion decay mode is into gaugino pairs. This decay mode will lead to additional
decay-produced LSP dark matter in the early universe.
The next four most dominant saxion decays modes are into $f\bar{f}$
(where $f$ stands for the various SM fermions),
Higgs plus gauge bosons, gauge boson pairs  and Higgs boson pairs.
All these modes can result in entropy dilution of any relics present
at the time of saxion decay. Saxion decay to sfermion pairs, which would also
add to decay produced LSPs, is highly suppressed.

\begin{figure}[tbh!]
    \centering
    \includegraphics[scale=0.48]{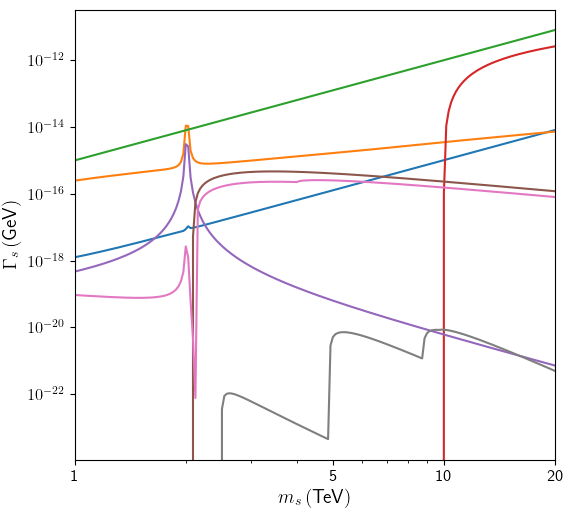}
    \includegraphics[scale=0.48]{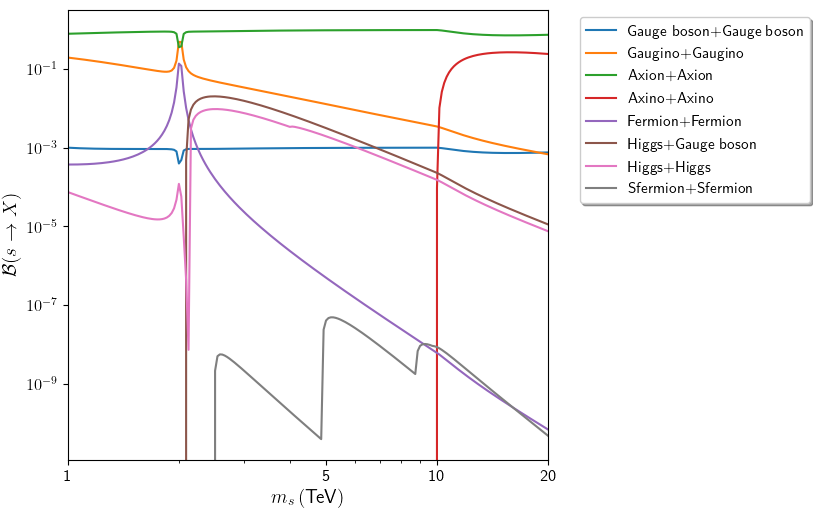}
    \caption{Saxion $s$  {\it a}) decay widths and
    {\it b}) branching fractions into PQMSSM particles  vs. $m_s$ for $\xi=1$.
    \label{fig:sBFxi1}}
\end{figure}

In Fig. \ref{fig:sBFxi1}, we again show saxion decay widths and branching
fractions, this time for $\xi=1$.
For this case, the dominant saxion decay mode is to axion pairs $s\to aa$,
leading to DR production in the early universe. Also, for
$m_s>2m_{\ta}$, then saxion decay to axino pairs $s\to\ta\ta$ turns on
and can be comparable to the saxion decay rate to $aa$. The saxion decay to axino pairs, followed by axino cascade decays, will lead to additional decay-produced LSP dark matter at the time of axino decay.
The remaining SM decay modes lead to entropy dilution.
For brevity, we do not show the various axino decay modes.
Under $R$-parity conservation, these modes are all into
particle+sparticle pairs.
These are displayed in Ref. \cite{Bae:2013hma} for a similar BM model
(which is now LHC excluded due to too low a gluino mass).

\subsection{Gravitino branching fractions}
\label{ssec:GBFs}

\begin{figure}[tbh!]
    \centering
    \includegraphics[scale=0.48]{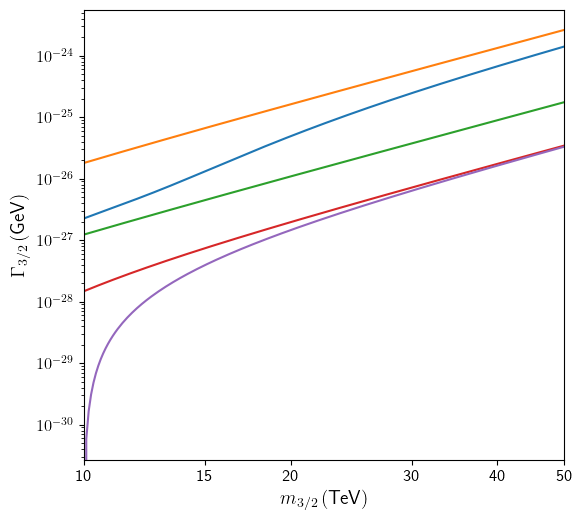}
    \includegraphics[scale=0.48]{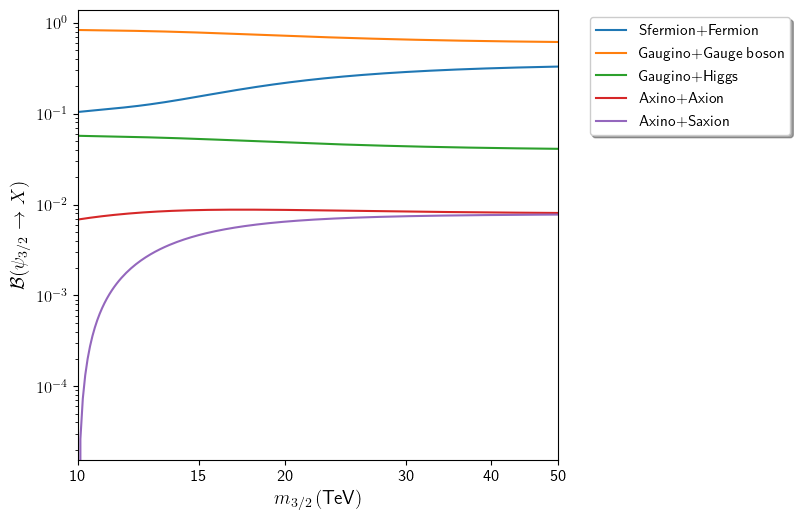}
    \caption{Gravitino $\psi_\mu$ {\it a}) decay widths and
    {\it b}) branching fractions into PQMSSM particles  vs. $m_{3/2}$.
    \label{fig:GBF}}
\end{figure}

For completeness, we show in Fig. \ref{fig:GBF} the gravitino decay widths and
branching fractions for our natural SUSY BM model versus
gravitino mass $m_{3/2}$.
These widths are programmed from formulae in
Kohri {\it et al.}\cite{Kohri:2005wn}.
All modes are into particle+sparticle pairs so can feed into decay-produced
LSP dark matter at the time of gravitino decay, but can also disrupt successful
BBN. The dominant decay mode is $\psi_\mu \to$ gauge boson+gaugino, followed by decay to fermion+sfermion, then gaugino+Higgs.
The gravitino decay branching fractions into PQ states $a\ta$ and $s\ta$ are
typically below the 1\% level.

\section{Nine coupled Boltzmann equations in the $\phi$PQMSSM model}
\label{sec:Boltz}

In this Section, we outline briefly our nine-coupled-Boltzmann equation
evaluation of mixed axion-WIMP dark matter production in the $\phi$PQMSSM model.
In Ref. \cite{Bae:2014rfa}, we had evaluated the relic abundance in the
PQMSSM (DFSZ case) with eight coupled Boltzmann equations.
As was discussed in Ref. \cite{Bae:2014rfa}, we use the Boltzmann equations in Eqs.\ref{eq:boltzmannNumberDensity} and \ref{eq:boltzmannEnergyDensity} for the thermally-produced/decay-produced
(TP/DP) components, which are 
1. TP/DP axions,
2. TP/DP saxions,
3. TP/DP axinos,
4. TP/DP gravitinos, and
5. TP/DP neutralinos.
The Boltzmann equation governing the evolution of number density for a species $i$ is given by
\begin{multline}
    \label{eq:boltzmannNumberDensity}
    \frac{
        dn_i
    }{
        dt
    }
    +
    3 
    H
    n_i
    =
    \sum_{j \in \text{MSSM}}
    \left(
        \overline{n}_i
        \overline{n}_j
        -
        n_i
        n_j
    \right)
    \langle 
    \sigma
    v
    \rangle_{ij}
    -
    \Gamma_i
    \frac{
        m_i n_i
    }{
        \rho_i
    }
    \left(
        n_i
        -
        \overline{n}_i
        \sum 
        \limits_{i \rightarrow a + b}
        \mathcal{B}_{i \rightarrow a + b}
        \frac{
            n_a 
            n_b
        }{
            \overline{n}_a
            \overline{n}_b
        }
    \right)
    \\
    +
    \sum_a
    \Gamma_a
    \frac{
        m_a
        n_a
    }{
        \rho_a
    }
    \left(
        \mathcal{B}_{a \rightarrow i}
        n_a
        -
        \overline{n}_a
        \sum 
        \limits_{a \rightarrow i + b}
        \mathcal{B}_{a \rightarrow i + b}
        \frac{
            n_i 
            n_b
        }{
            \overline{n}_i
            \overline{n}_b
        }
    \right)
\end{multline}
while the Boltzmann equation governing evolution of the energy density of $i$ reads
\begin{multline}
    \label{eq:boltzmannEnergyDensity}
    \frac{
        d\rho_i
    }{
        dt
    }
    +
    3 
    H
    \left(
        \rho_i
        +
        \mathcal{P}_i
    \right)
    =
    \sum_{j \in \text{MSSM}}
    \left(
        \overline{n}_i
        \overline{n}_j
        -
        n_i
        n_j
    \right)
    \langle 
    \sigma 
    v
    \rangle_{ij}
    \frac{
        \rho_i
    }{
        n_i
    }
    -
    \Gamma_i
    m_i
    \left(
        n_i
        -
        \overline{n}_i
        \sum 
        \limits_{i \rightarrow a + b}
        \mathcal{B}_{i \rightarrow a + b}
        \frac{
            n_a 
            n_b
        }{
            \overline{n}_a
            \overline{n}_b
        }
    \right)
    \\
    +
    \sum_a
    \Gamma_a
    \frac{
        m_a
    }{
        2
    }
    \left(
        \mathcal{B}_{a \rightarrow i}
        n_a
        -
        \overline{n}_a
        \sum 
        \limits_{a \rightarrow i + b}
        \mathcal{B}_{a \rightarrow i + b}
        \frac{
            n_i 
            n_b
        }{
            \overline{n}_i
            \overline{n}_b
        }
    \right)
    .
\end{multline}
Here, we denote ($\overline{n}_i$)$n_i$ as the (equilibrium) number density, $\rho_i$ the energy density, $\Gamma_i$ the total decay width, and $\mathcal{B}$ is the appropriate branching fraction denoted in the subscript.
Additionally, the Hubble parameter is given by $H = \sqrt{ \rho_{tot} / 3 m_P^2 }$.
The collision operators relevant for our case lead to the terms on the right-hand side, which are the familiar annihilation, decay, and injection terms - which run over all $a$ that decay to $i$.
The factor $m_i n_i / \rho_i$ present in the decay and injection terms serves as a relativistic dilation factor, which as we will see has important consequences for LSP DM production from the decay-produced saxions.
As was discussed in Ref. \cite{Bae:2014rfa} the decay and injection terms also account for inverse decays, which can be important in the DFSZ scenario and may prolong the decay of the saxion and axino.

These Boltzmann equations are also valid for the coherent oscillation (CO) modes upon setting 
the annihilation cross section $\langle \sigma v \rangle^{CO} = 0$, 
the pressure $\mathcal{P}_i^{CO} = 0$, 
the equilibrium number density $\overline{n}_i^{CO} = 0$,
and taking the injection terms to 0.
Additionally, for CO fields we always have $\rho_i = m_i n_i$.
With these modifications, Eqs.~(\ref{eq:boltzmannNumberDensity}) and (\ref{eq:boltzmannEnergyDensity}) also describe
6. the CO modulus,
7. the CO saxion, and 
8. the CO axion - which due to its temperature-dependent mass, requires the addition of the term $\dot{m} / m$ to Eq.~(\ref{eq:boltzmannEnergyDensity}) (see {\it e.g.} Ref. \cite{Kolb:1990vq}).
In this work, we again adopt the axion temperature-dependence as described in Ref. \cite{Visinelli:2009kt}.

Finally, we must close these equations with a ninth Boltzmann equation describing radiation.
Following Ref. \cite{Bae:2014rfa}, we adopt the following equation that models the evolution of entropy $S$:
\begin{equation}
    \label{eq:boltzmannEntropy}
    \frac{
        dS
    }{
        dt
    }
    =
    \frac{R^3}{T}
    \sum 
    \limits_i
    \mathcal{B}_{i \rightarrow \text{rad}}
    \Gamma_i
    m_i
    \left(
        n_i
        -
        \overline{n}_i
        \sum 
        \limits_{i \rightarrow a + b}
        \mathcal{B}_{i \rightarrow a + b}
        \frac{
            n_a 
            n_b
        }{
            \overline{n}_a
            \overline{n}_b
        }
    \right)
    .
\end{equation}

We have also upgraded the computer code of Ref. \cite{Bae:2014rfa} to a modular
form using the C\texttt{++} language. This allows us to utilize the peer-reviewed Boost library \cite{BoostLibrary}
for all special functions such as the Bessel functions which are used in calculation of cross sections.
These algorithms provide far more accurate results in the high-temperature regime than the power series approximations previously used (although much of the late-time cosmology is relatively unaffected so long as species reach equilibrium).
Additionally, we use Boost's Odeint library to numerically integrate the above Boltzmann equations, adopting the Rosenbrock 4 algorithm as the annihilation terms are numerically stiff.
This codebase was also designed to incorporate the semi-quantitative estimates used in {\it e.g.} Refs. \cite{Bae:2022okh,Baer:2022fou} to provide cross-checks for our results.
We intend on making this code publicly available in the near future.

Our calculation for the relic abundance begins by stipulating the initial
abundances at $T=T_R$ and $R=R_0$ where we must also include initial values for
the CO-produced saxion, axion and modulus fields $s_i$,
$a_i$ and $\phi_0$ where $\theta_i=a_i/f_a$. 
Our methodology follows that already described in Ref's. \cite{Baer:2011uz} and \cite{Bae:2013qr}
for the SUSY KSVZ axion model and in
Ref. \cite{Bae:2014rfa} for the SUSY DFSZ axion model.
The input parameters to the code are thus:
\be
T_R,\ f_a,\ \theta_i,\ s_i,\ m_{3/2},\ m_{\ta},\ m_s,\ m_\phi\ {\rm and}\ \phi_0 
\ee
and a SUSY BM point (and the modulus $\lambda_i$ couplings).
We examine two string-inspired cases: 1. the heterotic/M-theory inspired case \textbf{GK1} with $\lambda_{gauge}=1$
and 2. the Type IIB inspired case \textbf{GK2} with $\lambda_{gauge}=1/16\pi^2$.
In both cases, we fix the parameters 
$m_\phi = 1.8 \times 10^4$ TeV, 
$f_a =10^{11}$ GeV, 
$m_s = m_{\tilde{a}} = 5$ TeV, 
$s_i = f_a$, 
$\theta_i = 3.113$, 
and take the PQ self-coupling $\xi=1$ for the remainder of this section.
We also fix $\phi_0 = \sqrt{2/3} m_P$ which, as was remarked in Ref. \cite{Bae:2022okh}, is the maximum value of $\phi_0$ that is consistent with a radiation-dominated universe at $T=T_R$, unless $T_R \gtrsim 10^{12}$ GeV at which point Ref. \cite{Buchmuller:2004xr} argues the dilaton becomes destabilized.
Here, we take $T_R = 10^{10}$ GeV.
A version of the Isajet code IsaReD\cite{Baer:2002fv}, which was modified to use the Boost Bessel function algorithms \cite{BoostLibrary},
computes the thermally averaged neutralino (co)-annihilation times
relative velocity function $\langle\sigma v (T)\rangle$
needed for the neutralino Boltzmann equation.
The modification was required to obtain accurate values above $T \gtrsim 20$ GeV, which is a regime not required in IsaReD's relic density estimate.

\begin{figure}[htb!]
    \centering
    \includegraphics[height=0.3\textheight]{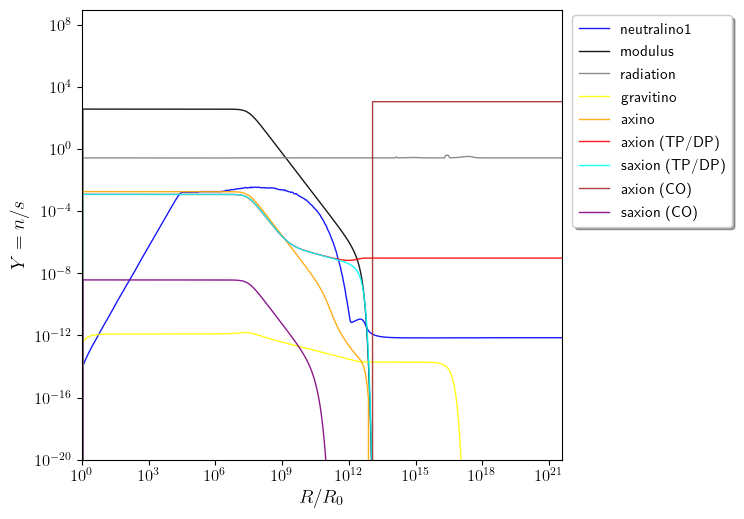}\\
    \includegraphics[height=0.3\textheight]{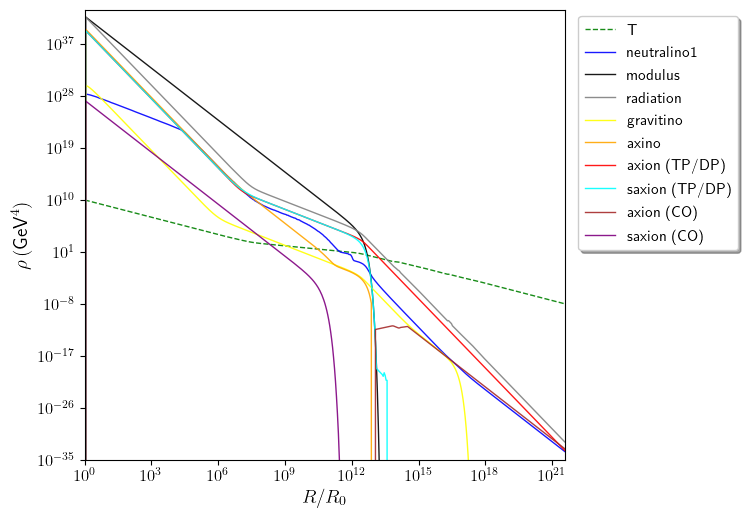}
    \caption{In {\it a}), we show the yield variable $Y=n/s$ vs. scale factor 
        $R/R_0$ from the time of inflationary reheat to the era of entropy conservation, for various matter
        and energy densities in the early universe, for our BM scenario in the PQMSSM/\textbf{GK1} 
        case. In {\it b}), we show evolution of the nine energy densities $\rho_i$ vs.
        scale factor $R$ from its value $R_0$ at inflationary reheat to the era of
        entropy conservation. 
        We also show the associated temperature of
        radiation $T$ where the scale for $T$ is listed on the left-hand-side
        instead in GeV units.
        For the figure, we adopt the natural SUSY BM point with
        $m_{\ta} =m_s =5$ TeV with $f_a=s_i=10^{11}$ GeV and
        $\phi_0 = \sqrt{2/3} m_P$ and $m_{\phi} =1.8 \times 10^4$ TeV and
        $T_R=10^{10}$ GeV with $\theta_i =3.113$ and $\xi =1$.
        In this {\it heterotic/M-theory} inspired case, all $\lambda_i$ couplings are equal to 1.
    \label{fig:RhoVsR_het}}
\end{figure}

Our first result is shown in Fig. \ref{fig:RhoVsR_het} for the case \textbf{GK1}
where we plot 
in frame {\it a}) the yield variables $Y=n_i/s$ and in {\it b}) the
nine various energy densities vs. early universe scale factor $R/R_0$
from the end of inflation to the era of entropy conservation.\footnote{
    Here, we use the notation $R_0$ to refer to the value of the scale factor at inflationary reheating - not to be confused with 
    much of the cosmological literature that uses $R_0 \equiv 1$ to refer to the present value.
}
Once we are in the era of entropy conservation, then $n_i/s$ is conserved
henceforth and the relic density can be computed as
\be
\Omega_ih^2=m_i(n_i/s)s_0 h^2/\rho_c
\ee
where $\rho_c \simeq 8.0992 h^2 \times 10^{-47} \text{ GeV}^4$ is the critical closure density, $h$ is the scaled Hubble constant, and where $s_0$ is the present-day entropy density of the universe:
$s_0=\frac{2\pi^2}{45}g_{*S}T_0^3 \simeq 2969.5 T_0^3 \text{ cm}^{-3}$.

From Fig. \ref{fig:RhoVsR_het}, we see that immediately at the end of
inflation where $T=T_R$ and $R/R_0=1$, the universe is momentarily radiation dominated (gray curve) with some presence of TP saxions, gravitinos, axions,
axinos and neutralinos. The modulus oscillation temperature
$T_{osc}$ was already reached during the inflationary reheating process, 
and the modulus field has already begun to oscillate behaving as CDM.
The universe very quickly becomes modulus dominated\cite{Giblin:2017wlo}
and stays that way until the
modulus field begins to decay around $R/R_0\sim 10^{13}$ (black curve).
The CO saxion field (purple curve) has also started to oscillate during inflationary reheating
and decays around $R/R_0\sim 10^{11}$, which augments the neutralino abundance.
The neutralino abundance begins near zero and soon builds up to 
its thermal equilibrium value. It is also augmented by the axino, TP/DP saxion, and CO modulus decay around
$R/R_0\sim 10^{12-13}$. 
The axino begins to decay around $R/R_0 \sim 10^{11}$, but then becomes sourced by the modulus decay, so that it stays present until around $R/R_0 \sim 10^{13}$ when the modulus decays. 
A similar effect occurs to the TP/DP saxion, although this is not noticeable in the plot as they are unsuppressed decays of the modulus, which washes out the influence of the saxion decay term until it is no longer sourced around $R/R_0\sim 10^{13}$, noticeably later than the CO-produced saxions.
Gravitinos decay around $R/R_0\sim 10^{17}$ which also feeds into the
neutralino abundance, although its contribution is only at roughly the $1\%-5\%$ percent level in our case, which assumes helicity-suppressed modulus decays to gravitinos.
Axions begin to oscillate around $R/R_0\sim 10^{13}$ and behave as CDM
and so end up dominating the energy abundance of the universe shortly after
$R/R_0\sim 10^{21}$.
By the era of entropy conservation, after all unstable
particles have decayed, only radiation, axions (TP/DP and CO-produced)
and neutralinos remain.
The TP/DP axions, which were augmented by modulus and saxion decay,
diminish as $1/R^4$ and so become subdominant, but contribute to
dark radiation $\Delta N_{eff}$.
The temperature of the universe vs. $R/R_0$ is also denoted in the Figure
by the green dashed line, where its scale is denoted on the
right-vertical axis but now in GeV units.
It decreases uniformly as $R$ increases, but less steeply around $R/R_0\sim 10^{7}-10^{12}$ 
where it is augmented by the
entropy injection from various unstable constituents, although predominantly due to decay of the modulus \cite{Scherrer:1984fd}.

\begin{figure}[tbh!]
    \centering
    \includegraphics[height=0.3\textheight]{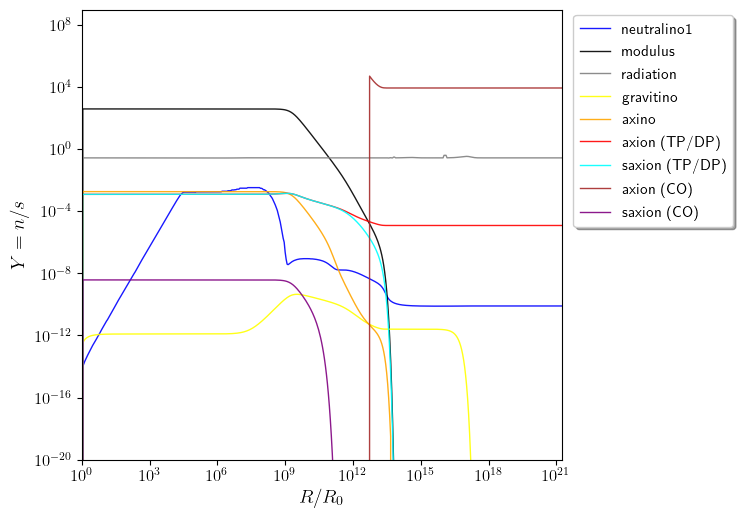}\\
    \includegraphics[height=0.3\textheight]{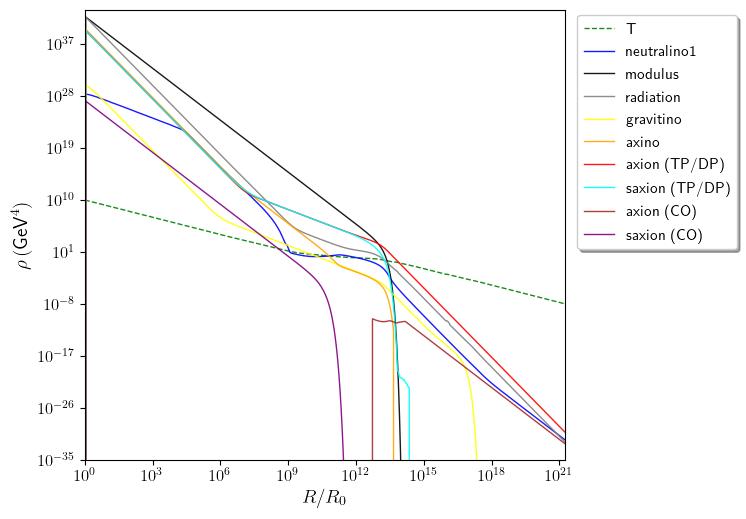}
    \caption{In {\it a}), we show the yield variable $Y=n/s$ vs. scale factor 
        $R/R_0$ from the time of inflationary reheat to the era of entropy conservation, for various matter
        and energy densities in the early universe, for case \textbf{GK2} in our BM scenario of the 
        $\phi$PQMSSM model. 
        In {\it b}), we show evolution of the nine energy densities $\rho_i$ vs.
        scale factor $R$ from its value $R_0$ at re-heat to the era of
        entropy conservation. 
        We also show the associated temperature of
        radiation $T$ where the scale for $T$ is listed on the left-hand-side
        instead in GeV units.
        For the figure, we adopt the natural SUSY BM point with
        $m_{\ta} =m_s =5$ TeV with $f_a=s_i=10^{11}$ GeV and
        $\phi_0 = \sqrt{2/3} m_P$ and $m_{\phi} =1.8 \times 10^4$ TeV and
        $T_R=10^{10}$ GeV with $\theta_i =3.113$ and $\xi =1$.
        In this {\it IIB} inspired case, most $\lambda_i$ couplings are equal to 1 except 
        $\lambda_{gauge}=1/16\pi^2$.
    \label{fig:RhoVsR_IIB}}
\end{figure}

In Fig. \ref{fig:RhoVsR_IIB}, we show in frame {\it a}) the yield variable and
in frame {\it b}) the various energy densities as a function of $R/R_0$ but this time
for the \textbf{GK2} BM scenario. In the \textbf{GK2} case, there is less entropy dilution from $\phi$ decay
than in the \textbf{GK1} case since now $\phi$ decay to $VV$ is suppressed. 
Also, in the \textbf{GK2} case, we see a greater augmentation of the neutralino abundance from $\phi$ 
decay since the gauge modes are suppressed in the \textbf{GK2} case (which are the leading modes in the \textbf{GK1} case).
This slightly reduces the decay temperature of the modulus, resulting in less efficient neutralino annihilations and an increase in their abundance.
To a lesser extent, the increase in branching fraction to SUSY modes - which is higher than in the \textbf{GK1} case - makes a small additional contribution to the neutralino abundance.
However, this increase is mostly washed out due to the neutralino annihilations that take place after they are produced, making the decay temperature of the modulus the dominant factor in neutralino relic abundance.
The slight change in the decay scale of the modulus can also be seen from the plot - which pushes modulus decay (and therefore the decays of the DP saxions and axinos) towards $R/R_0 \sim 10^{14}$.
Modulus decay also begins to overlap with the onset of axion oscillations, slightly reducing their abundance.

\begin{figure}[tbh!]
    \centering
    \includegraphics[height=0.35\textheight]{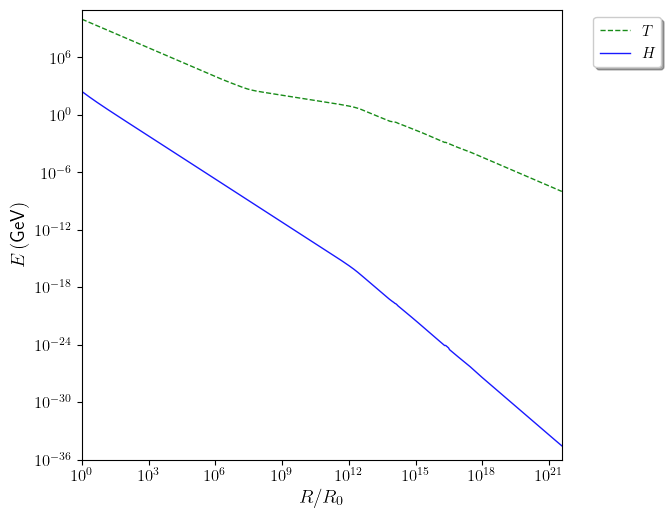}
      \caption{Evolution of the Hubble constant $H$ vs.
        scale factor $R$ from its value $R_0$ at re-heat to the era of
        entropy conservation. We also show the associated temperature of
        radiation $T$.
        This figure is created with the same data from Fig. \ref{fig:RhoVsR_het}.
    \label{fig:H}}
\end{figure}

In Fig. \ref{fig:H}, we show the magnitude of the Hubble constant $H$ in GeV units 
vs. $R/R_0$ for the same parameters as in Fig. \ref{fig:RhoVsR_het}.
The value of $H$ decreases as $R^{3/2}$ early on when the universe is
modulus dominated, but when $R/R_0\sim 10^{12}$, it decreases as $R^{2}$
once the $\phi$ field has decayed and the universe is no longer matter dominated.
We also show the corresponding temperature $T$ of radiation, denoted by the green dashed curve.

\section{Dark matter and dark radiation production in the $\phi$PQMSSM model}
\label{sec:Oh2_phiPQ}

In this Section, we examine rates for DM and DR production in the
$\phi$PQMSSM model using our nine-coupled-Boltzmann equation code.
The amount of dark radiation is given by\cite{Bae:2013qr}
\be
\Delta N_{eff}=\rho_a/\rho_\nu=\frac{120}{7\pi^2}\left(\frac{11}{4}\right)^{4/3}\frac{\rho_a}{T^4}
\ee
where $\rho_\nu$ is the energy density from a single species of of neutrino with
$\rho_\nu=\frac{7}{8}\frac{\pi^2}{15}T_\nu^4$ and $\rho_a$ is the energy
density in relativistic axions which contains contributions from
1. thermally produced axions\cite{Bae:2011iw} and
2. decay-produced axions where
the DP axions arise from gravitino, saxion and modulus decay.
We do not consider the presence of an ultralight ALP here.

\subsection{Decoupling solution to the CMP in the $\phi$PQMSSM}
\label{ssec:decoupling}

\subsubsection{Case \textbf{GK1}}

\begin{figure}[tbh!]
    \centering
    \includegraphics[height=0.3\textheight]{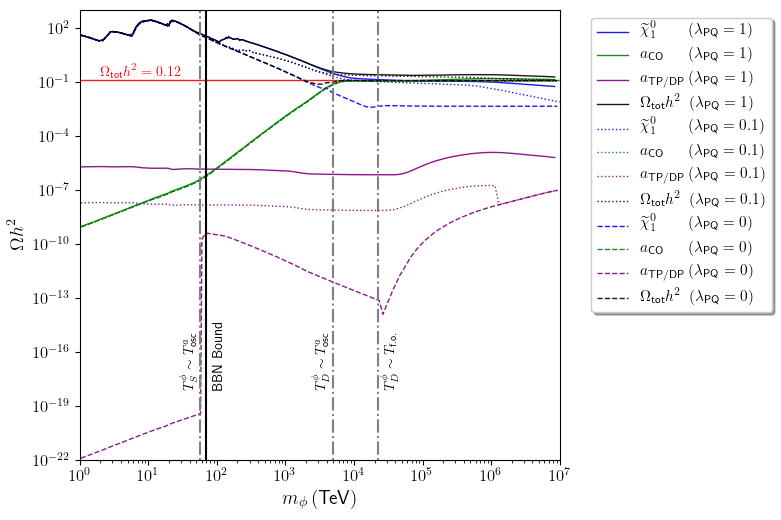}\\
    \includegraphics[height=0.3\textheight]{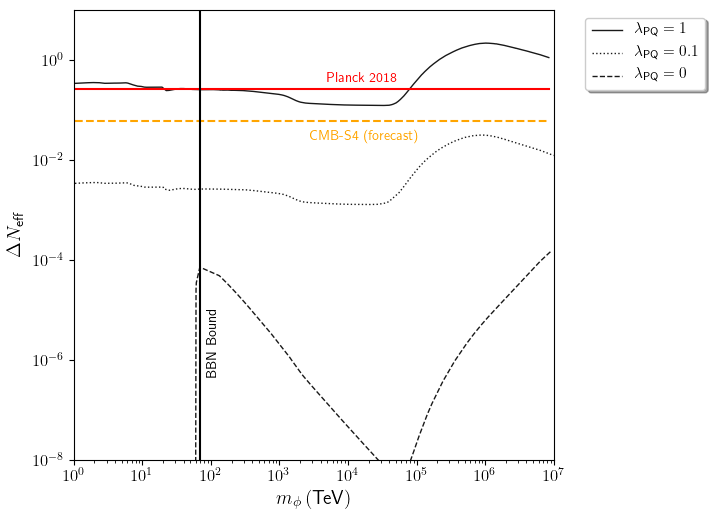}
    \caption{
        In {\it a}), we plot the values of various contributions to
        $\Omega h^2$
        (dark matter relic density) as generated
        vs. modulus mass $m_\phi$ from the $\phi$PQMSSM model with
        $\lambda_{gauge}=1$ (\textbf{GK1}) and $\lambda_{PQ}=0$ (dashed),
        $0.1$ (dotted) and 1 (solid) with $\xi =0$.
        The region left of the solid black vertical line is where modulus $\phi$
        decay violates BBN bounds.
        We also show the locus of various key temperatures where the value
        of $m_\phi$ corresponds to a key shift in the resulting cosmology.
        For the figure, we adopt the natural SUSY BM point with
        $m_{\ta}=m_s=5 $ TeV, $m_{3/2}=30$ TeV with $f_a=s_i=10^{11}$, $\phi_0=\sqrt{2/3}m_P$
        and $\theta_i=3.113$.
        In {\it b}), we plot the value of $\Delta N_{eff}$ (dark radiation)
        as generated from the
        $\phi$PQMSSM model with $\lambda_{PQ}=0$, $0.1$ and 1.
        We also show the present 95\% CL bounds
        on $\Delta N_{eff}$ from Planck 2018 results (red) and
        projected sensitivity of CMB-S4 (orange).
    \label{fig:xi0}}
\end{figure}

In Fig. \ref{fig:xi0}, we show in frame {\it a}) the relic densities of
1. neutralinos $\tchi_1^0$ (blue), TP/DP- (purple) and CO-produced (green) axions 
and the total DM abundance (black) as functions of the modulus mass
$m_{\phi}$ for $\lambda_{gauge}=1$ and for 
values of $\lambda_{PQ} \in \{ 0, 0.1, 1 \}$ and with $\xi =0$.
The remaining modulus couplings are fixed to $\lambda_i = 1$.
We also take $f_a=10^{11}$ GeV with $m_{\ta}=m_s=5$ TeV, $m_{3/2}=30$ TeV
and $\phi_0= \sqrt{2/3} m_P$ with $s_i=f_a$ and $\theta_i=3.113$.
Additionally, we plot in frame \textit{a}) of Fig. \ref{fig:xi0} three additional vertical dot-dashed lines which indicate changes in the resulting cosmology.
The first vertical dot-dashed line, located at $m_\phi \sim 50$ TeV, indicates the value of $m_\phi$ that begins to inject entropy into the thermal bath at the same time as the axion oscillations begin.
As can be seen from the green curve, to the right of this line the axion relic abundance increases more quickly.
This is because the CO axion is only diluted once oscillations begin and does not feel the full effect of the entropy dilution.
The second vertical dot-dashed line, located at $m_\phi \sim 5 \times 10^3$ TeV, then indicates where the decay temperature of the modulus and the oscillation of the axion coincide.
To the right of this line, we see that the axion now is at a constant relic density - as it no longer receives any dilution from the modulus decay.
The third vertical dot-dashed line is located at $m_\phi \sim 2\times 10^4$ TeV and indicates where the modulus decay temperature matches the neutralino freeze-out temperature.
At low values of $m_\phi$, we see
the neutralinos dominate the relic abundance and indeed that neutralinos
are extremely {\it overproduced} due mainly to $\phi\to SUSY$ particle
decay at low temperatures $T \ll T_{f.o.}$, where $T_{f.o.}\sim m_{\tchi}/20$ is the
neutralino freeze-out temperature. This is the case of modulus-induced
dark matter overproduction.
We also see at $m_\phi \sim 70$ TeV, the vertical dashed black line separates the BBN-violating region ({\it i.e.} $m_\phi \lesssim 70$ TeV, which is thus ruled out) from the BBN-safe region.
In this case, the ruled-out region persists for values of $m_\phi$ above
the BBN bound due to overproduction of WIMPs (even though for our benchmark model the WIMPs are thermally-underproduced). 
One must have $m_{\phi}$ as high as
  $\sim 10^4$ TeV in order for $\Omega_{\tchi}h^2$ to drop below the measured
$0.12$ value.
The neutralino abundance is enhanced from its TP value even for $m_\phi\agt 10^4$ TeV; this is because the dominant $\phi\to ss$ decay has a large cascade decay of $s\to SUSY$, 
while the saxions are highly relativistic ($m_\phi \gg 2 m_s$) leading to a large dilation of the saxion lifetime.
Here, we stress that even if the modulus decays before neutralino freeze-out, the DP saxions typically do \textit{not}.
Additionally, the neutralino abundance is enhanced compared to the $\xi=1$ case (next Figure) since the $s\to SUSY$ is enhanced in the $\xi=0$ case, while the absence of the $s \rightarrow \text{PQ}$ modes increases its lifetime.
The CO-produced axion abundance is quite small for small values of
$m_\phi$ due to substantial entropy dumping by the late modulus decay
which dilutes the abundance of all relics at the time of modulus decay.
The entropy dilution factor is given by $r\equiv S_f/S_0\simeq 4m_\phi Y_\phi/2T_D=T_e/T_D$ where $S_F$ is the final entropy density and $S_0$ is the initial
entropy density, $Y_\phi\equiv n_\phi/s$ is the modulus yield variable, $T_D$
is the temperature at which the modulus decays $T_D\sim \sqrt{\Gamma_\phi m_P}$
and $T_e\sim \sqrt{m_\phi \phi_0}$ is the temperature at which the
modulus energy density equals the radiation density. As $m_{\phi}$ increases,
$r$ decreases (less entropy dilution due to an earlier modulus decay) and
the CO-produced axion energy density increases.
Once $m_{\phi}$ increases beyond the dotted line
where $T_D\simeq T_{osc}^a$, then the modulus decays before the axion field
starts oscillating and there is no further dilution. 
Meanwhile, the TP/DP axion abundance (purple curves) is always quite low, and does not contribute significantly to the overall DM abundance. 
Thus, for $m_{\phi}\agt 10^4$ TeV,
we see that the measured DM relic density is attained. 
The lesson learned here is that
substantially larger values of $m_\phi$ are needed to solve the moduli-induced
dark matter overproduction problem as opposed to the moduli-induced BBN
problem!

\begin{figure}[tbh!]
    \centering
    \includegraphics[height=0.3\textheight]{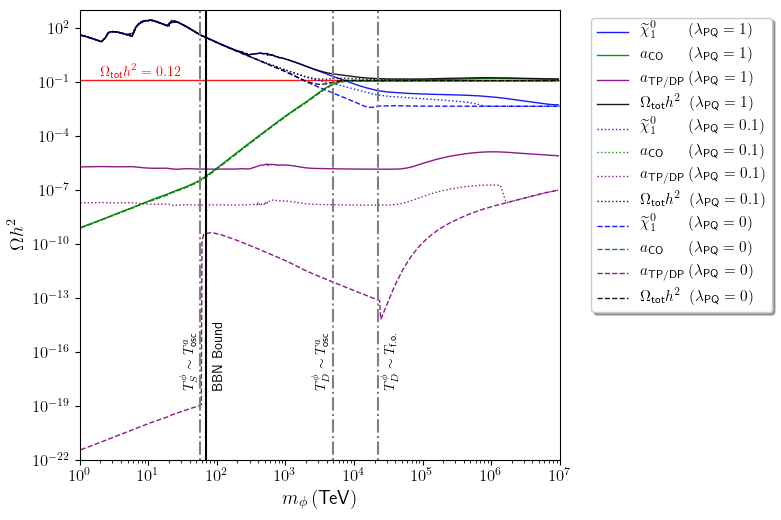}\\
    \includegraphics[height=0.3\textheight]{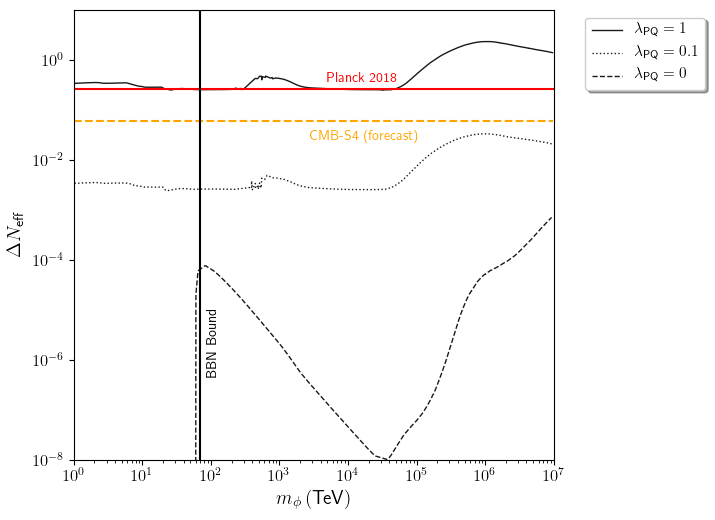}
    \caption{
        In {\it a}), we plot the values of various contributions to
        $\Omega h^2$
        (dark matter relic density) as generated
        vs. modulus mass $m_\phi$ from the $\phi$PQMSSM model with
        $\lambda_{gauge}=1$ (\textbf{GK1}) and $\lambda_{PQ}=0$ (dashed),
        $0.1$ (dotted) and 1 (solid) with $\xi =1$.
        The region left of the solid black vertical line is where modulus $\phi$
        decay violates BBN bounds.
        We also show the locus of various key temperatures where the value
        of $m_\phi$ corresponds to a key shift in the resulting cosmology.
        For the figure, we adopt the natural SUSY BM point with
        $m_{\ta}=m_s=5 $ TeV, $m_{3/2}=30$ TeV with $f_a=s_i=10^{11}$, $\phi_0=\sqrt{2/3}m_P$
        and $\theta_i=3.113$.
        In {\it b}), we plot the value of $\Delta N_{eff}$ (dark radiation)
        as generated from the
        $\phi$PQMSSM model with $\lambda_{PQ}=0$, $0.1$ and 1.
        We also show the present 95\% CL bounds
        on $\Delta N_{eff}$ from Planck 2018 results (red) and
        projected sensitivity of CMB-S4 (orange).
    \label{fig:xi1}}
\end{figure}

In frame {\it b}), we show the associated dark radiation $\Delta N_{eff}$
where the horizontal red line denotes the Planck 2018 limit. We see that for
$m_\phi\alt 10^2$ TeV (in the BBN excluded region) then DR is slightly
overproduced due to late-time modulus decay.
Also, for $m_{\phi}\agt 10^5$ TeV with $\lambda_{PQ}\sim 1$, then DR is
again overproduced. This situation is explained as follows.
If the axion is produced {\it late} (to the left of the freeze-out
temperature $T_{f.o.}$ with $m_\phi\alt 1.5\times 10^4$ TeV), then the
axion production cross section is too low to have any big effect - the axions do not thermalize at all.
However, if the axion is produced earlier ($m_\phi\agt 1.5\times 10^4$ TeV),
the produced number density is below the equilibrium number density,
but the cross section becomes large enough to draw the axion population
towards equilibrium, resulting in an increase in both $\Omega_a^{TP}h^2$ and
$\Delta N_{eff}$.  
But, if DP-axions are produced {\it too early}
(when the curve starts falling for $m_{\phi}\agt 10^6$ TeV),
it gets enough closer to the equilibrium distribution that it begins to
behave like a matter distribution (based on the pressure term, which
determines the equation of state based on the ratio $\rho / n$
in comparison to $m$) and begins to get redshifted.  
However, for the $\lambda_{PQ}=0$ curve (bottom dashed), the story is actually
related to the entropy dilution factor which does begin to decrease
(by $m_{\phi}\sim 10^6$ TeV, entropy dilution is only $\sim 10^5$ as
compared to $\sim 10^{11}$ or so for lower $m_{\phi}$).
The result is that for $\lambda_{PQ}=1$, there exists a rather narrow window
$m_\phi:10^4-10^5$ TeV where neither DM nor DR is overproduced. For
smaller values of $\lambda_{PQ}$, then the DR is suppressed
and the model is allowed for the higher values of $m_\phi >10^5$ TeV as well.

In Figure \ref{fig:xi1}, we again show the various $\Omega_ih^2$ values
in frame {\it a}) and $\Delta N_{eff}$ values in frames {\it b})
for the same parameters as in Fig. \ref{fig:xi0} except that now
$\xi =1$ so that the decay $s\to aa$ can proceed ($s\to \ta\ta$ is
phase space forbidden) which both lowers the $\phi\to ss\to SUSY$
branching fraction and decreases the saxion lifetime. The main difference in the two figures is then that
the neutralino abundance for $m_\phi\agt 10^4$ TeV is somewhat suppressed
for $\xi=1$ with $\lambda_{PQ}\ne 0$ as compared to the $\xi =0$ case from
Fig. \ref{fig:xi0}, while the dark radiation production is increased slightly from the $s \rightarrow aa$ decays.
In the $\xi=1$ case, it appears that Planck 2018 results \cite{Planck:2018vyg} rule out $\lambda_{PQ} \gtrsim 1$.
However, $\lambda_{PQ} = 0.1$ is still below the forecast CMB-S4 limits \cite{Abazajian:2019tiv} (orange dashed line), which will probe much of the parameter space of expected dark radiation production for this scenario.

\subsubsection{Case \textbf{GK2}}
\label{ssec:xi1GK2}

\begin{figure}[tbh!]
    \centering
    \includegraphics[height=0.3\textheight]{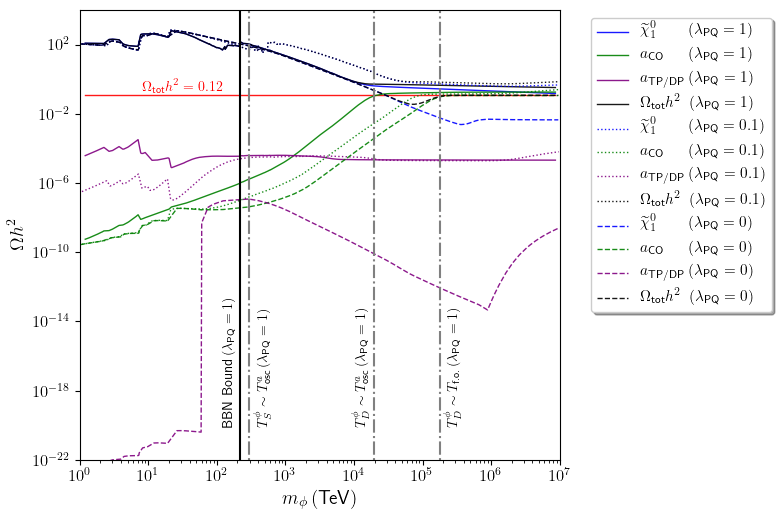}\\
    \includegraphics[height=0.3\textheight]{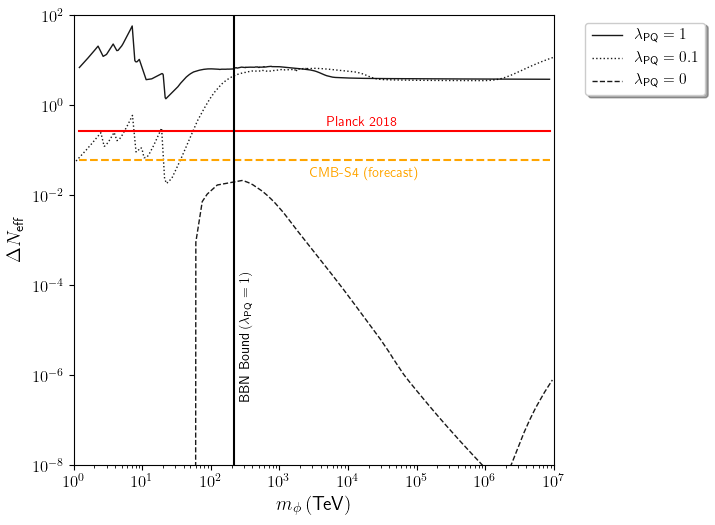}
    \caption{
        In {\it a}), we plot the values of various contributions to
        $\Omega h^2$
        (dark matter relic density) as generated
        vs. modulus mass $m_\phi$ from the $\phi$PQMSSM model with
        $\lambda_{gauge}=1/16\pi^2$ (\textbf{GK2}) and $\lambda_{PQ}=0$ (dashed),
        $0.1$ (dotted) and 1 (solid) with $\xi =0$.
        The region left of the solid black vertical line is where modulus $\phi$
        decay violates BBN bounds.
        We also show the locus of various key temperatures where the value
        of $m_\phi$ corresponds to a key shift in the resulting cosmology.
        For the figure, we adopt the natural SUSY BM point with
        $m_{\ta}=m_s=5 $ TeV, $m_{3/2}=30$ TeV with $f_a=s_i=10^{11}$, $\phi_0=\sqrt{2/3}m_P$
        and $\theta_i=3.113$.
        In {\it b}), we plot the value of $\Delta N_{eff}$ (dark radiation)
        as generated from the
        $\phi$PQMSSM model with $\lambda_{PQ}=0$, $0.1$ and 1.
        We also show the present 95\% CL bounds
        on $\Delta N_{eff}$ from Planck 2018 results (red) and
        projected sensitivity of CMB-S4 (orange).
    \label{fig:xi0_IIB}}
\end{figure}

In Fig. \ref{fig:xi0_IIB}{\it a}), we show the relic abundances of various $\phi$PQMSSM
constituents vs. modulus mass $m_\phi$ for the Type-IIB inspired case \textbf{GK2} with
a suppressed value of $\lambda_{gauge}=1/16\pi^2$ but with PQ self-coupling $\xi =0$, 
and other parameters as in previous benchmark figures. 
Since here the dominant decay of the modulus is to the PQ sector, the decay scale of the modulus is directly tied to the value of $\lambda_{PQ}$ as all other modes are suppressed in some form.
For the lower range of $m_\phi\alt 200$ TeV, the modulus decays after BBN starts for $\lambda_{PQ} = 1$. 
The other vertical lines indicating cosmology shifts are also pushed to larger $m_\phi$ for $\lambda_{PQ}=1$ than in the previous case.
In these plots, we display the vertical line for only $\lambda_{PQ} = 1$, and all four vertical lines get pushed to higher $m_\phi$ for lower values of $\lambda_{PQ}$.
For values of $m_\phi \gtrsim 200$ TeV, then the \textbf{GK2} case is BBN safe but neutralino dark matter is still grossly overproduced due to late modulus decay to SUSY particles.
For $10^3 \text{ TeV} \lesssim m_\phi \lesssim 10^4 \text{ TeV}$, we see that the neutralino DM abundance for $\lambda_{PQ} = 1$ and $\lambda_{PQ} = 0$ overlap, while $\lambda_{PQ} = 0.1$ has a higher neutralino abundance.
This is a direct consequence of the sensitive dependence on
$\lambda_{PQ}$: for $\lambda_{PQ} = 0.1$, the modulus decay occurs sooner than the $\lambda_{PQ} = 0$ case.
However, the DP saxions are very relativistic in this regime, which leads to a late saxion decay almost exclusively into gauginos thus producing a large enhancement in the abundance of neutralinos.
For $\lambda_{PQ} = 1$, the dilation of the saxion lifetime is unchanged, but the decay temperature of the modulus is increased by a full order of magnitude, resulting in an order of magnitude decrease in the neutralino density.
The fact that these two curves overlap in this region is coincidental in this case.
Even for large enough $m_\phi\agt 10^5$ TeV where $\phi$ decays before neutralino freeze-out $T_{f.o.}$
(dot-dashed vertical line), neutralinos are still overproduced as long as $\lambda_{PQ}$ is not too small. 
For $\lambda_{PQ}\sim 0$, then neutralinos can assume their thermally-frozen out value, 
where we have underproduced higgsinos (dashed blue line). 
For $m_\phi \gg 10^7$ TeV, eventually the modulus decay will occur early enough to offset the relativistic dilation of the saxion, which would finally reduce the neutralino abundance to its thermal value.
Also, for $m_\phi\alt 10^4$ TeV, then
moduli decay after the onset of axion oscillations, thus diluting the CO-produced axions.

In Fig. \ref{fig:xi0_IIB}{\it b}), we show the associated contribution to dark radiation, $\Delta N_{eff}$, which is almost exclusively from decay-produced axions. 
The Planck bound on $\Delta N_{eff}$ is shown by the red horizontal line. 
For case \textbf{GK2} which has suppressed $\lambda_{gauge}$, the modulus dominantly decays to
saxions and axions as shown in Fig. \ref{fig:phiBF2}. 
As we have taken $\xi=0$ here, the saxions do not contribute to $\Delta N_{eff}$ - with the dark radiation instead saturated entirely by the axions produced directly from modulus decay.
This leads to gross overproduction of dark radiation, thus excluding this entire scenario unless
$\lambda_{PQ}\sim 0$ (thus turning off  the modulus coupling to the PQ sector).  
Such a scenario would either require fine-tuning of $\lambda_{PQ}$ or additional symmetries which we do not consider here.

\begin{figure}[tbh!]
    \centering
    \includegraphics[height=0.3\textheight]{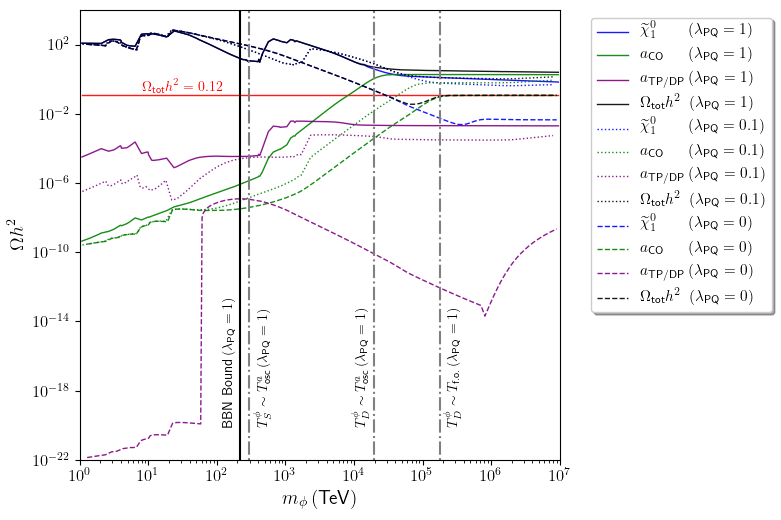}\\
    \includegraphics[height=0.3\textheight]{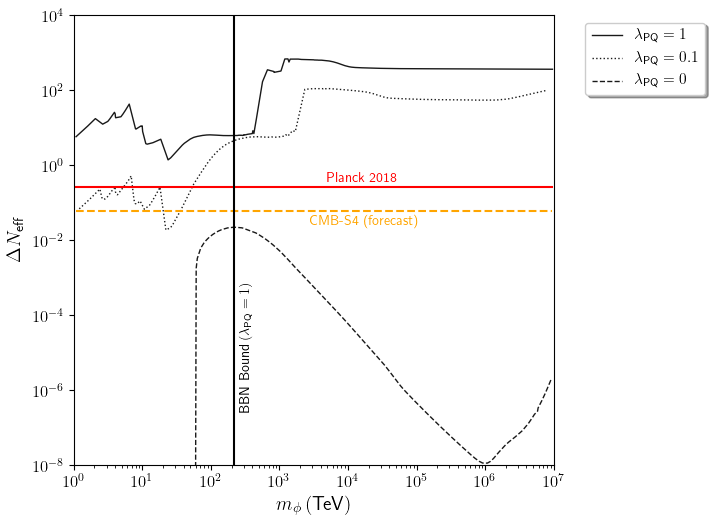}
    \caption{
        In {\it a}), we plot the values of various contributions to
        $\Omega h^2$
        (dark matter relic density) as generated
        vs. modulus mass $m_\phi$ from the $\phi$PQMSSM model with
        $\lambda_{gauge}=1/16\pi^2$ (\textbf{GK2}) and $\lambda_{PQ}=0$ (dashed),
        $0.1$ (dotted) and 1 (solid) with $\xi =1$.
        The region left of the solid black vertical line is where modulus $\phi$
        decay violates BBN bounds.
        We also show the locus of various key temperatures where the value
        of $m_\phi$ corresponds to a key shift in the resulting cosmology.
        For the figure, we adopt the natural SUSY BM point with
        $m_{\ta}=m_s=5 $ TeV, $m_{3/2}=30$ TeV with $f_a=s_i=10^{11}$, $\phi_0=\sqrt{2/3}m_P$
        and $\theta_i=3.113$.
        In {\it b}), we plot the value of $\Delta N_{eff}$ (dark radiation)
        as generated from the
        $\phi$PQMSSM model with $\lambda_{PQ}=0$, $0.1$ and 1.
        We also show the present 95\% CL bounds
        on $\Delta N_{eff}$ from Planck 2018 results (red) and
        projected sensitivity of CMB-S4 (orange).
    \label{fig:xi1_IIB}}
\end{figure}

In Fig. \ref{fig:xi1_IIB}{\it a}), we show the similar case of the IIB-inspired string scenario \textbf{GK2}
but with PQ self coupling factor $\xi=1$ (which allows for $s\to a a$ decays). In this case, 
neutralino dark matter is overproduced across all values of $m_\phi$ up to and beyond
$m_\phi\sim 10^7$ TeV. 
As in the $\xi=0$ case we just studied, even as $T_D^\phi$ exceeds $T_{f.o.}$, the modulus
decay produces huge amounts of saxions, which decay much later into SUSY particles, thus augmenting the relic abundance. 
In this case, CO-produced axions are also enhanced.
This enhancement occurs significantly in only this case because here, most of the energy of the modulus is transferred to dark radiation instead of returning to the thermal bath.
As we saw in Fig. \ref{fig:sBFxi1}, for $\xi=1$ the saxion decays primarily into axions - and since the modulus decays almost exclusively to axions and saxions in this case, a large majority of all cascade decays of the modulus end in axions.
Since the DP axions will not thermalize, the radiation temperature is decreased when the CO axions begin to oscillate - in this scenario $T_{osc}^a \sim 0.6-0.7$ GeV, while for all previous cases $T_{osc}^a \sim 1$ GeV.
This small decrease in the axion oscillation temperature thus corresponds to a large increase in the initial axion energy density since $m_a \propto T^{-4}$ above $T \sim \Lambda_{QCD} \sim 200$ MeV and $\rho_a^0 = \frac{1}{2} m^2(T_{osc}^a) a_i^2$.

In Fig. \ref{fig:xi1_IIB}{\it b}), we show the associated dark radiation $\Delta N_{eff}$.
Here, we see DR is again overproduced across the range of $m_{\phi}$ values, thus excluding this scenario (even more than the $\xi=0$ case). 
As we have just discussed, here the $s\to aa$ decays are turned on with $\xi =1$, 
and so $\phi\to ss$ decay followed by $s\to aa$ decay amplifies the total
DR which is produced.  Specifically, for most of the parameter space, the 
effective branching ratio of the modulus to dark radiation is above around $95\%$ for this case.

\subsubsection{Entropy dilution}

\begin{figure}[tbh!]
    \centering
    \includegraphics[height=0.35\textheight]{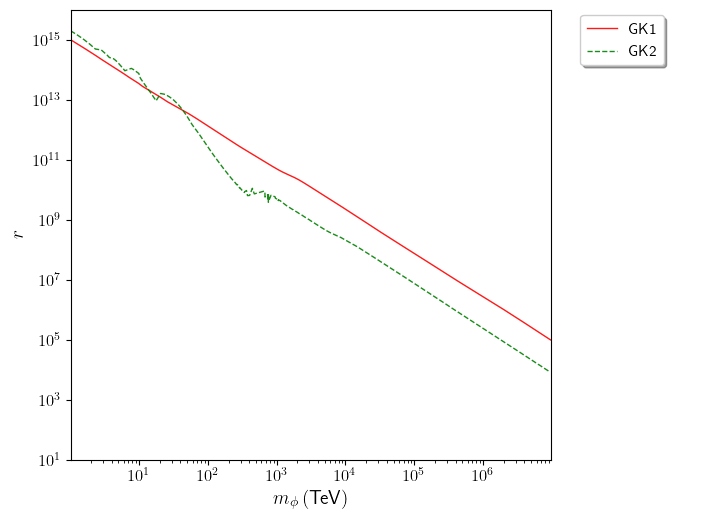}
    \caption{Entropy dilution factor $r$ versus modulus mass $m_\phi$.
        For the figure, we adopt the natural SUSY BM point with
        $m_{\ta} =m_s =5$ TeV with $f_a=s_i=10^{11}$ GeV and
        $\phi_0 = \sqrt{2/3} m_P$  and
        $T_R=10^{10}$ GeV with $\theta_i =3.113$ and $\xi =1$.
    \label{fig:r}}
\end{figure}

In Fig. \ref{fig:r}, we show the entropy dilution factor $r$ vs. $m_{\phi}$
for the two cases \textbf{GK1} and \textbf{GK2}.
The entropy dilution is especially enormous $\sim 10^{15}$ for low values
of $m_{\phi}\sim 1$ TeV  where the modulus field decays at very late times,
within the BBN era.
For $m_{\phi}$ as high as $\sim 10^7$ TeV, then $r$ drops as low
as $\sim 10^5$.
The case \textbf{GK1} curve is rather smooth since the $\phi$ decay is dominated by
decay to gauge bosons; for the case \textbf{GK2} where these modes are suppressed,
then the curve is more dependent on the onset of various $\phi$ decay
modes into sparticles.
Additionally, this figure agrees well with Fig. 16 of Ref. \cite{Bae:2022okh}, which was created using semi-quantitative methods, while Fig. \ref{fig:r} in this work displays the entropy dilution computed from numerical solutions of the Boltzmann equations.
For extremely massive moduli, {\it e.g.} $m_\phi \gtrsim 10^7$ TeV, near total dilution of thermal relics may no longer be possible.
This may translate into a resurgence of the thermal gravitino problem \cite{Kawasaki:2008qe,Kawasaki:2017bqm} if $m_\phi$ is too large.

\subsubsection{Inflationary reheating temperature}

\begin{figure}[tbh!]
    \centering
    \includegraphics[height=0.35\textheight]{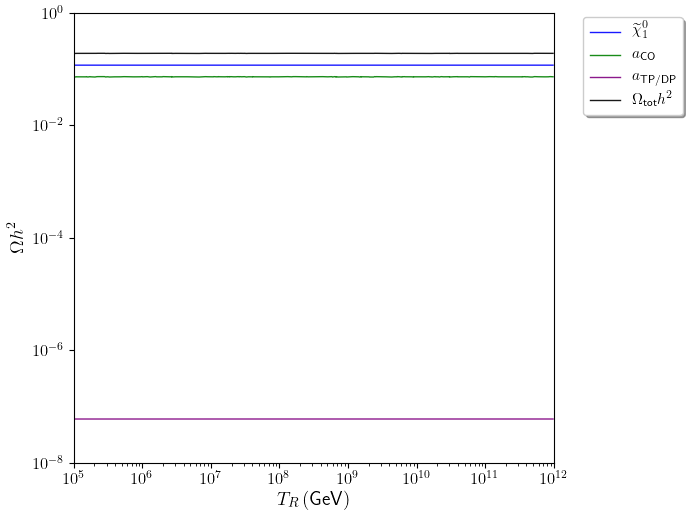}
    \caption{Inflationary reheat temperature $T_R$ versus
        relic abundance $\Omega_i h^2$.
        For the figure, we adopt the natural SUSY BM point in case \textbf{GK1} with
        $m_{\ta} =m_s =5$ TeV with $f_a=s_i=10^{11}$ GeV and
        $\phi_0 =\sqrt{2/3} m_P$ and $m_{\phi} =5 \times 10^3$ TeV with $\theta_i =3.113$ and $\xi =1$.
        Here, we take $\lambda_{PQ} = 0.2$.
        All other $\lambda_i$ couplings are equal to 1.
    \label{fig:reheatTemp}}
\end{figure}

One of the primary consequences of a modulus-dominated cosmology is the large amount of entropy injected to the thermal bath, diluting all previous relics and effectively resetting the initial conditions for the matter content of the universe.
The temperature of inflationary reheating is then not expected to change the late-time abundances.
This expectation agrees with our findings in Fig. \ref{fig:reheatTemp}
Within the scanned range, $T_R \in [ 10^5 \text{ GeV}, \, 10^{12} \text{ GeV} ]$, we see that the value of $T_R$ has effectively no consequence whatsoever on the produced relic abundances.
We find the same result for $\Delta N_{eff}$, although we do not display this plot for brevity.
As we discussed in the last section, one possible exception to this might be if $m_\phi \gtrsim 10^7$ TeV, when the entropy dilution becomes small.
However, we do not consider this case here.

\subsubsection{Surveying the PQ parameter space for $\xi=1$}

Finally, we investigate how these results are dependent on the PQ sector parameters.
Other than the masses of the saxion and axino, the parameters we are primarily interested in here are the axion initial misalignment angle $\theta_i$ and the PQ scale $f_a$.
Conventionally, the expected range of $f_a$ is $10^9 \text{ GeV} \lesssim f_a \lesssim 10^{12} \text{ GeV}$, where the lower bound comes from supernova cooling \cite{Brockway:1996yr,Grifols:1996id} and the upper bound is where axions typically overproduce the DM abundance without significant tuning of $\theta_i$.

\begin{figure}[tbh!]
    \centering
    \includegraphics[height=0.4\textheight]{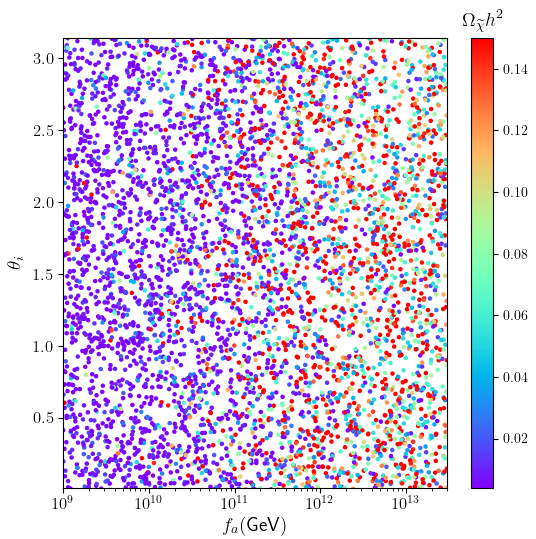}
    \caption{
        Neutralino DM abundance from scanning $f_a, \, \theta_i, \, m_s, \, m_{\tilde{a}}, \, \lambda_i$ for $m_\phi = 5 \times 10^{5} $ TeV.
        Red points have $\Omega_{\tchi} h^2 \geq \Omega_{meas} h^2 \sim 0.12$, while purple points are close to the thermal value.
        Additionally, purple and dark blue points can satisfy DD/ID constraints, while all other colors violate these bounds.
        We adopt the same natural SUSY BM point and case \textbf{GK1} with $\xi=1$.
    \label{fig:wimpDMfavstheta}}
\end{figure}
\begin{figure}[tbh!]
    \centering
    \includegraphics[height=0.4\textheight]{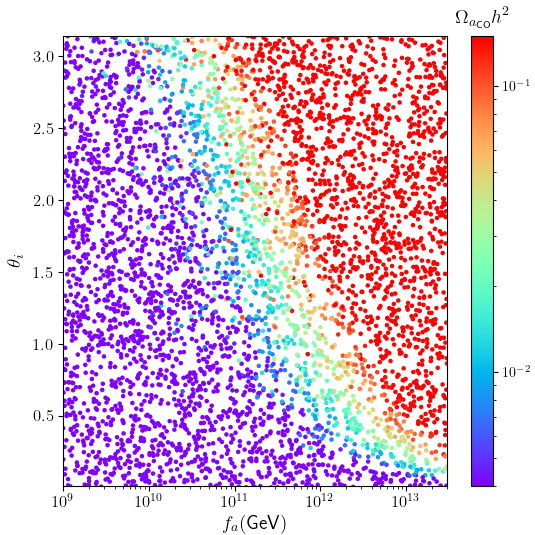}
    \caption{
        CO-produced axion DM abundance from scanning $f_a, \, \theta_i, \, m_s, \, m_{\tilde{a}}, \, \lambda_i$ for $m_\phi = 5 \times 10^{5} $ TeV.
        Red points have $\Omega_{a_{CO}} h^2 \geq \Omega_{meas} h^2 \sim 0.12$.
        Note we use a logarithmic color scale in this plot.
        We adopt the same natural SUSY BM point and case \textbf{GK1} with $\xi=1$.
    \label{fig:coAxionDMfavstheta}}
\end{figure}

In Fig. \ref{fig:wimpDMfavstheta}, we display the produced neutralino DM abundance in the $f_a$ vs $\theta_i$ plane.
Here, we take our natural SUSY benchmark point in case \textbf{GK1} with $\xi=1$ and fix $m_\phi = 5\times 10^{5}$ TeV.
We scan over $f_a \in [ 10^9 \text{ GeV}, 3 \times 10^{13} \text{ GeV} ]$ and $\theta_i \in [ 0, \pi ]$, and also scan over 
$m_s,\, m_{\tilde{a}} \in [ 1 \text{ TeV}, 30 \text{ TeV} ]$.
Additionally, we scan over modulus couplings $\lambda_i \in [ 0.1, 10 ]$  (including $\lambda_{gauge}$) in accordance to what one may expect for the NUHM3 model:
couplings to the first and second generation are randomly set to a unified value, while the third generation is set randomly to another value within our $\lambda_i$ interval, etc.
Scanning over the many unknown parameters throughout their expected ranges in this model, we find we are able to draw some rather general conclusions on this model for a given $m_\phi$ despite ignorance of the underlying parameter set.
We see that: quite generally neutralinos become overproduced as $f_a$ increases, with red points oversaturating the measured DM abundance.
This result is expected in that, for the $\xi=1$ case, the saxion lifetime depends on $f_a$ - increasing $f_a$ will increase the saxion lifetime.
Thus, when the saxions decay the produced neutralinos annihilate less efficiently than if they were produced at a higher temperature, resulting in an increase in their abundance.
However, due to the many parameters present here, for larger values of $f_a$ it is still possible to produce neutralino abundances that are not only below the measured $\Omega_{meas} h^2 \sim 0.12$, but also meet direct detection (DD) and indirect detection (ID) constraints.
Based on Ref. \cite{Baer:2016ucr}, we expect that for our natural SUSY BM point which is listed in Table 2 of Ref. \cite{Bae:2022okh}, if the neutralino DM abundance is less than around $10\%$ of the measured value, {\it i.e.} $\Omega_{\tchi} h^2 \lesssim 0.1 \Omega_{meas}h^2 \sim 0.012$, we satisfy bounds set by the XENON1T \cite{XENON:2018voc} experiment.\footnote{
    We should note that this BM point runs into mild tension with recent LUX-ZEPLIN results \cite{LZ:2022ufs}, however we do not expect use of a new natural SUSY BM point that satisfies this bound to change our results here in any significant way.
}
The purple points and the very dark blue points of Fig. \ref{fig:wimpDMfavstheta} are then expected to satisfy these DD/ID constraints.

We plot the produced CO axion abundance in Fig. \ref{fig:coAxionDMfavstheta} using the same scan results as in the previous paragraph.
Here, the relic abundance of these axions is very sensitive to $f_a$ and $\theta_i$ - but largely insensitive to all other parameters (as we have fixed $m_\phi = 5 \times 10^5$ TeV here).
The red points again show CO axion relic abundance which is in excess of the measured value.
Due to the strong dependence on $f_a$ and $\theta_i$ here, we also use a log scale for the color coding.
We see a fairly predictive band between $f_a \sim 10^{11}$ GeV and $f_a \sim 10^{13}$ GeV for various $\theta_i$ that separates underproduced and overproduced abundances of CO axions.
There do exist some red points to the left of this band, however these points arise similar to what we saw in Sec.~(\ref{ssec:xi1GK2}) - here the $\lambda_{PQ}$ coupling is randomly assigned towards the top of the scan limit ($\sim 10$), while $\lambda_{gauge}$ is randomly assigned a value close to the lower limit ($\sim 0.1$).
Thus, most of the energy from the modulus goes into dark radiation - increasing the CO axion abundance but excluding these points from vast DR overproduction.
We also note that, for most of the parameter space for $f_a \lesssim 10^{11}$ GeV, both neutralinos and CO axions are underproduced - although severe tuning of $\theta_i = \pi$ may push $\Omega_{a_{CO}}h^2 \sim 0.12$ due to anharmonic effects \cite{Visinelli:2009kt}.

\begin{figure}[tbh!]
    \centering
    \includegraphics[height=0.4\textheight]{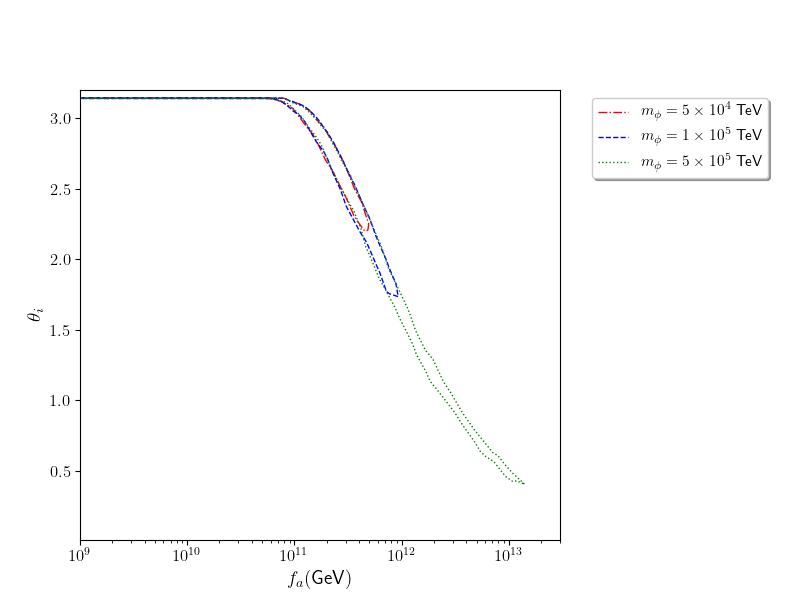}
    \caption{
        Allowed contours from scanning $f_a, \, \theta_i, \, m_s, \, m_{\tilde{a}}, \, \lambda_i$ for given values of $m_\phi$.
        Interior regions of contours saturate DM density $0.09 \leq \Omega_{tot} h^2 \leq 0.125$, have acceptable DR production $\Delta N_{eff} \leq 0.29$,
        and satisfy DD/ID constraints for the neutralino, $\Omega_{\tchi} h^2 \leq 0.1 \Omega_{meas} h^2$.
        We adopt the same natural SUSY BM point and case \textbf{GK1} with $\xi=1$.
    \label{fig:contours}}
\end{figure}

Additionally, we plot contours of allowed regions that saturate the DM abundance in Fig. \ref{fig:contours}.
Here, we fix values of $m_\phi \in \{ 5\times 10^4, \, 1\times 10^{5}, \, 5 \times 10^{5} \}$ TeV and scan over the same parameters within the same scan limits as before, with the exception that we artificially reduced our $f_a$ and $\theta_i$ bounds to surround the expected allowed region for the sake of increasing efficiency of allowed points.
In all cases, we have checked that the endpoints of our $f_a$ and $\theta_i$ regions were beyond the accepted values, and thus these contours are expected to be an accurate representation of allowed parameter space in this model.
We see that for $m_\phi = 5 \times 10^{4}$ TeV, shown by the red contour, $f_a \lesssim 5 \times 10^{11}$ GeV is allowed, while for larger values of $f_a$, neutralinos are produced beyond current DD/ID limits or even exceed $\Omega_{meas}h^2$.
As $m_\phi $ increases to $m_\phi = 1\times 10^{5}$ TeV (blue contour), the modulus decay occurs just enough earlier to allow $f_a \lesssim 10^{12}$ GeV, which also allows for lower values of $\theta_i$ to saturate the DM bound.
Finally, for $m_\phi = 5 \times 10^{5}$ TeV which is displayed by the green contour, we see that nearly the full interval $\theta_i \in [0, \pi]$ is allowed, pushing the maximum allowed $f_a$ to roughly $1-2 \times 10^{13}$ GeV.
For larger $m_\phi$ than is displayed on this plot, we expect very similar contours to $m_\phi = 5 \times 10^{5}$ TeV, where the neutralino abundance begins to relax towards its thermal value as seen from Fig. \ref{fig:xi1}.
For $m_\phi \lesssim 10^4$ TeV, we would not expect any region of this parameter space to both saturate the observed DM density while also satisfying current DR and DD/ID constraints.
By scanning over the modulus couplings between each $\lambda_i \in [0.1,10]$, we also find that no points with the ratio $\lambda_{PQ} / \lambda_{gauge} > 1$ are allowed, with smaller ratios of these two couplings being more likely to meet constraints.

Of course, the region to the below-left of our contours is not necessarily excluded - although this region severely \textit{underproduces} total DM.
Thus, one could argue that this region of parameter space allows for both inclusion of our $\phi$PQMSSM model, but the primary DM component would come from some other source (possibly other exotic stringy remnants or primordial black holes, for example).
We do not consider this case here - but simply note that this model can also be incorporated into other DM paradigms so long as the modulus is sufficiently heavy ($m_\phi \gtrsim 10^4$ TeV) in this region of parameter space.

\subsection{Modulus amplitude and the $\phi$PQMSSM}
\label{ssec:phi0}

The classic solution to the cosmological moduli problem, as shown above, is the decoupling solution
wherein $m_\phi\to\ {\rm large}$ values such that $\phi$ decays before the onset of BBN, 
and perhaps even before neutralino freeze-out $T_D>T_{f.o.}$. 
The decoupling solution typically requires $m_\phi\agt 10^3-10^4$ TeV 
to avoid overproduction of neutralino dark matter. 
In Ref. \cite{Baer:2021zbj}, an alternative solution to the CMP
was suggested, based upon cosmological (anthropic) selection of the
initial modulus amplitude $\phi_0$ even while $m_\phi\sim$ TeV-scale. 
It was found that a value of $\phi_0 \sim 10^{-7} m_P$ would be required to solve the moduli-induced LSP DM overproduction problem.
A $\phi_0$ this low is unlikely to be compatible with current understanding of inflation in string theory, which requires $\phi_0 / m_P \sim (0.1 - 1)$ without some enhanced symmetry \cite{Cicoli:2016olq,Das:2015uwa,Dine:1995uk,Dine:1995kz}.
Here, we do not worry whether or not this scenario can be actually be realized in a way consistent with the string landscape and inflationary paradigms;
although we study the dependence of our results on the modulus initial amplitude $\phi_0$, the value required to ascertain a viable cosmology may be beyond those realizable in explicit constructions \cite{Baumann:2014nda}.
 
\begin{figure}[tbh!]
    \centering
    \includegraphics[height=0.35\textheight]{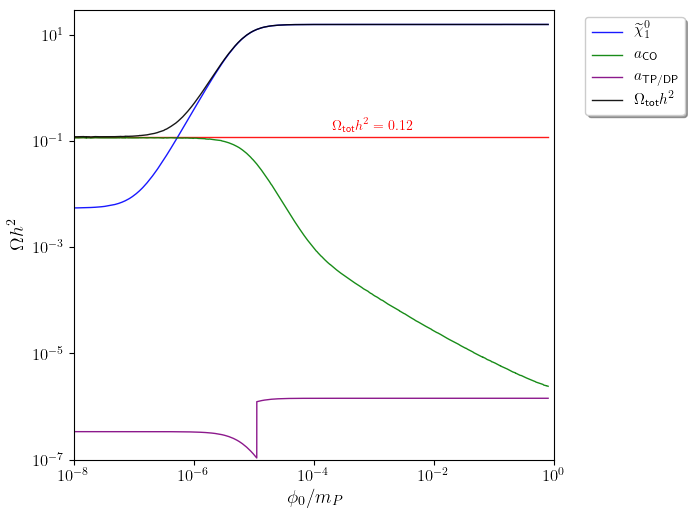}
    \caption{Dark matter relic densities for axions and neutralinos versus 
    modulus amplitude $\phi_0/m_P$ for
        our natural SUSY BM point in case \textbf{GK1} with
        $m_{\ta} =m_s =5$ TeV with $f_a=s_i=10^{11}$ GeV and
        $m_{\phi} =100$ TeV and $T_R=10^{8}$ GeV with $\theta_i =3.113$ and $\xi =1$.
        All $\lambda_i$ couplings are equal to 1.
    \label{fig:phi0}}
\end{figure}

In Fig. \ref{fig:phi0}, we show the neutralino and axion (and summed) relic densities
for our $\phi$PQMSSM BM point in case \textbf{GK1} with $\xi=1$, with $m_\phi = 100$ TeV, 
and versus $\phi_0/m_P$. Additionally, we have taken $T_R = 10^{8}$ GeV and all $\lambda_i = 1$.
It is usually assumed that $\phi_0\sim m_P$ in inflationary cosmology, and in this case we
would have an overabundance of neutralino dark matter. As $\phi_0$ is reduced to lower
values, then the neutralino relic density decreases into the $\Omega_{\tchi}h^2\sim 0.1$
range for $\phi_0\alt 10^{-5}m_P$. This is because the decreased modulus amplitude translates into 
a reduced modulus abundance, and hence much less neutralino production via moduli cascade decays. Here, we see the
axion relic density, which is initially underabundant, actually increases with
decreasing $\phi_0$ due to less entropy dilution from $\phi$ decay. 
For $\phi_0\alt 10^{-6}m_P$, then the relic abundance becomes axion-dominated and in accord
with its measured value in our universe. The TP/DP axion abundance assumes a tiny value. 
In the regime typically expected to be consistent with the string landscape, $0.1 \lesssim \phi_0/m_P \lesssim 1 $, we see that the only significant change is in CO axions which are still vastly underproduced.

\section{Summary and conclusions}
\label{sec:conclude}

We have examined dark matter and dark radiation production in the $\phi$PQMSSM
model wherein 1. SUSY provides a solution to the gauge hierarchy problem,
2. the SUSY DFSZ model provides a solution to the strong CP problem and
the SUSY $\mu$ problem wherein the PQ symmetry emerges as an accidental,
approximate global symmetry, perhaps  from some more fundamental discrete
${\bf Z}_n^R$ symmetry (such as ${\bf Z}_{24}^R$ as in Ref. \cite{Baer:2018avn}), and where
the axion quality problem is solved, and where the PQ scale $f_a$ is related
to the SUSY breaking scale $f_a\sim m_{SUSY}\sim 10^{11}$ GeV in the cosmological
sweet spot, and 3. we assume stringy unification with gravity giving rise
to a light, TeV-scale modulus field $\phi$ which is gravitationally coupled
to both the MSSM and the PQ sector. In the present case, this allows for
$\phi\to aa$, $ss$ and $\ta\ta$ decays which can heavily influence the
production of DM and DR in the early universe.
DM and DR production in the early universe is then sufficiently complicated
that its evaluation requires the solution of nine coupled Boltzmann equations
which track the various relic particles plus radiation which are assumed
to be present.

We find that DM and DR production in the $\phi$PQMSSM model has of course
a CMP, wherein the modulus field may
1. decay after BBN starts, thus disrupting the successful predictions of light element abundances in BBN,
2. decay to dark matter in the form of neutralinos and axions which
can be overproduced and
3. decay to DR can be overproduced since moduli, saxions and gravitinos
may all decay into a population of relativistic axions.
Assuming modulus coupling strengths $\lambda_i\sim 1$, then DM and/or DR are
generically overproduced for $m_\phi\alt 10^7$ GeV, and we must invoke the usual
modulus decoupling solution to the CMP wherein the modulus mass
$m_\phi$ must be taken so high that the modulus decay temperature
$T_D$ is not only above $T_{BBN}\sim 3-5$ MeV, but also $T_D\agt T_{f.o.}\sim m_{\tchi}/20\sim 10$ GeV (for natural SUSY models with $\mu\alt 350$ GeV).
Alternatively, if future work in string inflation can realize an anthropic solution to the CMP as was suggested in Ref. \cite{Baer:2021zbj},
the modulus mass could be taken to be $m_\phi\sim m_{soft}$ while $\phi_0/m_P \sim 10^{-7}$ would avoid a vast overproduction of dark matter, 
which would result in cosmic structures built mainly of DM with minimal baryons and likely be uninhabitable.
Large values for $m_\phi\agt 10^4$ TeV would be in conflict with
SUSY naturalness if the modulus mass $m_\phi$ is comparable to the masses of
MSSM scalars\cite{Bae:2022okh}. Alternatively, in some models
(such as {\it e.g.} the spectrum of scales anticipated in some KKLT constructions such as
Ref. \cite{Choi:2005ge}, wherein $m_{soft}\ll m_{3/2}\ll m_\phi$, or in sequestered models where $m_{soft} \ll m_\phi \ll m_{3/2}$ \cite{Aparicio:2014wxa,Blumenhagen:2009gk}) where
$m_\phi$ is allowed to be much greater than $m_{soft}$, then
naturalness and the CMP can be reconciled.

{\it Acknowledgements:} 

We thank L. Randall for suggesting this project many years ago.
This material is based upon work supported by the U.S. Department of Energy, 
Office of Science, Office of High Energy Physics under Award Number DE-SC-0009956 and U.S. Department of Energy (DoE) Grant DE-SC-0017647. 
The computing for this project was performed at the OU Supercomputing Center for Education \& Research (OSCER) at the University of Oklahoma (OU).


\bibliography{phipq}
\bibliographystyle{elsarticle-num}

\end{document}